\documentclass[12pt]{article}
\usepackage[english]{babel}
\font\Sets=msbm10

\def\Real{\hbox{\Sets R}}
\def\Complex{\hbox{\Sets C}}
\def\bb{\begin{equation}}
\def\ee{\end{equation}}
\def\sgn{\hbox{sgn}}
\def\Re{\hbox{Re}}
\def\Im{\hbox{Im}}
\def\pt{\partial}
\def\a{\alpha}
\def\b{\beta}

\def\k{\kappa}
\def\l{\lambda}
\def\o{\omega}
\def\O{\Omega}

\def\ve{\varepsilon}

\newtheorem{theorem}{Theorem}
\newtheorem{lemma}{Lemma}

\begin{document}

\title{Asymptotic behaviour of a solution for
Kadomtsev-Petviashvili-2 equation
\thanks{This work was partially supported by  RFBR (97-01-459),
grant for scientific schools (001596038) and INTAS (No1068).}}
\author{O.M. Kiselev\\Institute of Mathematics\\112 Chernyshevsky
str., Ufa, 450000, Russia\\ E-Mail: ok@imat.rb.ru}
\date{16th March 2000}
\maketitle

\abstract{An asymptotic behaviour of solution of
Kadomtsev-Petviashvili-2 equation is obtained as
$t\to\infty$ uniformly with respect to spatial
variables.}
\section{Introduction}

\par
Properties of solutions and process of integration for the
Kadomtsev-Petviashvili equation (KP) \cite{kp}: \bb \pt_x(\pt_t
u+6u\pt_x u+\pt^3_x u)=-3\sigma^2\pt^2_y u \label{kp} \ee depend on
sign of $\sigma^2$. This  equation is called KP-2 if $\sigma^2=1$.
\par
In this  work the asymptotic behaviour of decreasing
solution of equation KP-2 is obtained as $t\to\infty$.
The main term of the asymptotics has an order by
$O(t^{-1})$ and fast oscillates. An envelope of these
oscillations depends on $\xi=x/t$ and $\eta=y/t$.
Results of  the work are formulated in terms of
scattering data for auxiliary linear problem, which
is associated with equation KP-2 in the  inverse
scattering transform (IST) \cite{z-sh,dr}.
\par
The equation KP plays important role in modern mathematical physics.
Many  applications of these equation are known in  plasma physics,
water waves and other fields of waves process \cite{Petv-Pokh}.
Therefore questions about solvability of this equation in different
functional classes were studied in detail. In particular,  the
existence of global solution of Cauchy problem in class of
distribution functions was proved in \cite{Fam}. A norm of solution
in Sobolev space was estimated and an asymptotics of a solution were
obtained as $t\to\infty$ for KP-like equation but with more high
nonlinearity, which called generalized equation KP, in \cite{H-N}.
\par
The IST formalism allows to reduce a constructing of
the solution of nonlinear integrable equation into
solving of linear problems. One of most important
achievements of IST is the constructing of
asymptotic  behaviours of solutions as $t\to\infty$.
These results are well-known for 1+1-dimensional
equations (one spatial and  one temporal variable)
\cite{a-n}-\cite{Its}. Asymptotic behaviours of
solutions for (2+1)-dimensional equations were
studied not so in detail. Rigorous results  about
temporal asymptotics of solutions, which is
nonuniform with respect to spatial variables, for a
special class of nondecreasing solutions were
obtained in \cite{A-K-Kh}, \cite{O-P-Kh}. In work
\cite{M-S-T} an formal asymptotics of decreasing
solution of equation KP-1, which is  nonuniform with
respect to spatial variables, was constructed.
Results about asymptotic behaviour of decreasing
solutions  of equation  KP-2 as $t\to\infty$ and
about a remainder of this asymptotics,  which are
uniform with respect to spatial variables,  are
obtained in present work.
\par
The formalism IST for solving of Cauchy problem for
equation KP-2 was presented in \cite{A-Y-F}. This
formalism will be used in this work. Below we remind
basic steps of solving of Cauchy problem for equation
KP-2 by IST.
\par
Let us denote an initial condition for equation KP-2 as:
\begin{equation}
u|_{t=0}=u_0(x,y). \label{init}
\end{equation}
The process of solving of the Cauchy problem (\ref{kp}),
(\ref{init}) consists of several steps.
\par
First step is solving direct scattering problem. On this step  a
boundary problem is solved for function $\varphi$:\bb -\pt_y
\varphi+\pt^2_x \varphi+2ik\pt_x\varphi+u\varphi=0, \quad
\varphi|_{|k|\to\infty}=1,  \label{sche} \ee and  scattering data
are constructed by a formula:
\begin{eqnarray}
F(k) =& {\sgn(-\Re(k))\over2\pi}\int\int_{\Real^2}dx\,dy \,
u_0(x,y)\varphi(x,y,k,0) \times \nonumber\\ &\times \exp(-i(k+\bar
k)x-(k^2-\bar k^2)y). \label{sd}
\end{eqnarray}
It is useful to note, if $u$ is real, then the function $F(k)$ has
property $F(-\bar k)=-\bar F(k)$. This follows from formulas
(\ref{sche}) and (\ref{sd}).
\par
On the next step an evolution of scattering data is determined. This
evolution is very simple: $$ {\cal  F}(k;t)=F(k)\exp(4it(k^3+\bar
k^3)).$$
\par
Third step is solving of inverse scattering problem.  This is
reduced to so-called  $\bar D$-problem:\bb
\begin{array}{c}
\pt_{\bar k}\varphi=\psi\,F(-\bar k)\exp(itS),\\
\pt_k\psi=-\varphi\,F(k)\exp(-itS);
\end{array}
\left(\begin{array}{c}
\varphi\\
\psi
\end{array}\right)|_{|k|\to\infty}
= \left(\begin{array}{c} 1\\1
\end{array}\right),
\label{bar d} \ee where $S=4(k^3+\bar k^3)+(k+\bar k)\xi-i(k^2-\bar
k^2)\eta,\quad \xi=x/t,\,\eta=y/t$, the function $F(k)$ is
nonanalytic with respect to complex variable $k\in\Complex$. Solving
of this problem allows to obtain the functions $\varphi$ and $\psi$
at any time, if we know the evolution of scattering data.
\par
At last we can obtain the solution of the Cauchy problem by using
the formula \cite{A-Y-F}: \bb
u(x,y,t)=\pt_x\int\int_{\Complex}dk\wedge d\bar k\,
F(k)\psi(k,x,y,t) \exp(itS). \label{sol} \ee
\par
However, this successive method of solving for Cauchy problems for
general initial  conditions allows to obtain very implicit answer
with respect to initial data. Therefore the problem (\ref{bar d}) and
integral (\ref{sol}) are used usually  for solving of integrable
nonlinear equations and the scattering data is taking from some
suitable functional class (see, for example,
\cite{A-K-Kh}-\cite{M-S-T}). On one-hand side this gives for studying
a functional class of solutions, but on  the other hand-side this
leads to implicit restrictions on class of initial conditions.
\par
The auxiliary linear problem (\ref{sche}) and some related problems
were studied in  \cite{N-H}-\cite{Grin}. In the work \cite{Grin} it
was  proved that if the  function $u_0(x,y)$ decreases
exponentially, then  the  direct scattering problem (\ref{sche}) is
solvable for $\forall k\in\Complex$.
\par
To construct the asymptotics of $u$ as $t\to\infty$ we use the
asymptotics of $\psi$ and evaluate an asymptotics of the integral
(\ref{sol}) by stationary phase method \cite{Fedoryk}. As it turned
out, the main term of asymptotics of the function $u$ may be defined
by using only the main term of an asymptotic expansion of $\psi$ as
$t\to\infty$. Taking into account the fast oscillating coefficients
of the system from (\ref{bar d}) one can guess, that $\psi=1+o(1)$ as
$t\to\infty$. But we cannot say something definitely about the
asymptotic behaviour of the function $u$, until we do not know
analytic properties and an order of the remainder of asymptotics of
$\psi$  more precisely. Thus we come to studying an asymptotic
behaviour of a solution of the $\bar D$-problem (\ref{bar d}).
\par
The asymptotic behaviour of the solution of the $\bar D$-problem as
$t\to\infty$ with continuous  and fast oscillated coefficients was
obtained in \cite{ok}. Here we study the $\bar D$-problem with
discontinuous coefficients on imaginary axis of complex parameter
$k$. Constructing of the asymptotics of solution for such problem is
more complicated. Not only stationary points of phase function of an
oscillated exponent define a structure of the asymptotic solution
for (\ref{bar d}) as in \cite{ok} (see also \cite{ok1}), but the
location of these stationary points with respect to line of
discontinuity of coefficients of the equations (\ref{bar d}) as
well. It contributes additional difficulties into evaluating and
leads to changes in results. The uniform asymptotics of the solution
of the problem (\ref{bar d}) is constructed by matching method
\cite{I}.

\section{Main result}

\par
\begin{theorem}
Let $(1+|k|)F\in L_1\cap C$, $\pt^\a F\in L_1\cap C$
as $\Re(k)\not=0$, $|a|\le2$ and: \bb
\sup_{z\in\Complex}\int\int_{\Real^2}d\kappa d\lambda
\left\vert{F(\kappa+i\lambda) \over \kappa+i\lambda
-z} \right\vert\,<\,2\pi, \label{restriction} \ee
then the solution of the Cauchy problem of equation
KP-2 for corresponding initial condition exists as
$\forall t>0$. The asymptotic behaviour of the
solution as $t\to\infty$ differs in different domains
of variables $(x,y,t)$:
\newline
as $-(12\xi+\eta^2)t^{1/3}\gg1$:
\begin{eqnarray*}
u(x,y,t)=-4t^{-1}{\pi\over12i\sqrt{-\eta^2-12\xi}}
f\bigg({1\over2}\sqrt{-\eta^2-12\xi}+{i\eta\over12}\bigg)\times
\\
\times \exp\bigg(-11it\sqrt{-{y^2\over t^2}-12{x\over
t}}\bigg)+c.c.+o(1). \end{eqnarray*} as $(12\xi+\eta^2)t^{1/3}\gg1$:
$$ u=o(t^{-1}); $$ as $|12\xi+12\eta^2|\ll1$:
\begin{eqnarray*}
u(x,y,t)=8it^{-1}\sqrt{\pi}f(i\eta/12) \bigg( \int_0^\infty
dp_1\sqrt{p_1} \cos\bigg(8p_1^3-z p_1\bigg)+
\\+
\int_0^\infty dp_1\sqrt{p_1} \sin\bigg(8p_1^3-z p_1\bigg)\bigg)
\,+\,o(t^{-1}).
\end{eqnarray*}
Here $\xi=x/t$, $\eta=y/t$, $$ z=8\bigg({y^2\over12t^{4/3}}+{x\over
t^{1/3}}\bigg); $$
\begin{eqnarray*}
f(k) ={1\over2\pi}\int\int_{\Real^2}dx\,dy \,
u_0(x,y)\varphi(x,y,k,0)\times
\\
\times\exp(-i(k+\bar k)x-(k^2-\bar k^2)y).
\end{eqnarray*}
\end{theorem}
\par
The domains of validity for the asymptotics of the  solution of
equation KP-2 are intersected and therefore they give combined
asymptotics of the solution uniformly on plane of $x,y$.

\section{An  analytic behaviours of the scattering  data}

\par
In this section we demonstrate analytic behaviour of the scattering
data which  corresponds to sufficiently smooth and decreasing
initial condition (\ref{init}).
\par
First of  all we show, that the solution of (\ref{bar d}) exists.
\begin{theorem}
Let  $F(k)$ be such, that the condition
(\ref{restriction}) is fulfilled, then the solution
of (\ref{bar d}) exists in a space of continuous
vector-functions bounded when $k\in\Complex$.
\end{theorem}
\par
{\bf Proof.} Let us consider a system of integral equations which is
equivalent to (\ref{bar d}): $$ \left(\begin{array}{c}\varphi\\
\psi\end{array}\right)=
\left(\begin{array}{c}1\\
1\end{array}\right)+ G[F]\left(\begin{array}{c}\varphi\\
\psi\end{array}\right), $$ where
\begin{eqnarray*}
G[F]V= \int\int_{m\in\Complex}dm\wedge d\bar m\times\qquad\qquad
\qquad\qquad\qquad\qquad\qquad\\
\left(
\begin{array}{cc}
0 & {F(-\bar m)\over k-m}\exp(itS)\\ {F(m)\over
\overline{k-m}}\exp(-itS) & 0
\end{array}
\right)V(m,\xi,\eta,t).
\end{eqnarray*}
Using  well-known results   about integral operators (see, for
example \cite{Vlad}) one can show that the operator $G[F]$
transforms the space of continuous vector-functions into itself.
\par
The operator $G[F]$ is contracting operator. It is follows from
inequality (\ref{restriction}). Hence we obtain the theorem
statement.
\par
To construct the asymptotics we  must study smoothness of scattering
data $F(k)$ in neighborhood of line of  discontinuity $\Re(k)=0$.
\begin{lemma}
Let $u_0(x,y)$ be a finite function,  then in neighborhood of some
point $k'\in\Complex$ the scattering data have to be represented in
the form: $$ F(k)=\sgn[\Re(-k)]\sum_{|\alpha|=0}^{2}f_{\a_1\a_2}(k')
(k-k')^{\a_1}\overline{(k-k')^{\a_2}}\, +O(|k-k'|^3), $$ where
$f_{\a_1,\a_2}(k')$  are continuous functions, when $\Re(k')\not=0$
and $$f_{\a_1,\a_2}(k')=f_{\a_1\a_2}^{(1)}(k')+
\sgn[\Re(-k)]f_{\a_1\a_2}^{(2)}(k'), $$ when $\Re(k')=0$.
\end{lemma}
{\bf Proof} of this lemma consists of successive evaluating of
partial derivatives of scattering data  with respect to $k$ and
$\bar k$. For example let us evaluate $f_{10}^{(j)}(k')$ and
$f_{10}^{(j)}(k')$ as $\Re(k')=0$.
\par
Denote $$ f(k)={1\over2\pi}\int\int_{\Real^2}dx\,dy \,
u_0(x,y)\varphi(x,y,k,0)\exp(-i(k+\bar k)x-(k^2-\bar
k^2)y). $$ Then $f_{00}^{(1)}(k')=f(k')$,
$f_{00}^{(0)}(k')\equiv0$.
\par
Evaluate $f_{10}(k')=f_{10}^{(1)}(k')+\sgn[\Re(-k)]f_{10}^{(0)}(k')$.
\begin{eqnarray*}
\pt_k f(k)={1\over2\pi}\int\int_{\Real^2}dx dy \,
u_0(x,y)
\exp(-i(k+\bar k)x-(k^2-\bar k^2)y)\times\\
\big[(ix-2ky)\varphi(x,y,k,0)+\pt_k
\varphi(x,y,k,0)\big].
\end{eqnarray*}
To evaluate the derivative $\pt_k\varphi$ we use the integral
equation for the function $\varphi$:
\begin{eqnarray*}
\pt_k\varphi= {-1\over2i\pi}V.P.\int\int_{\Complex}{dm\wedge d\bar
m\over (k-m)^2} F(-\bar m)\times\\
\exp(i(m+\bar m)-(m^2-\bar m^2)y)\psi(x,y,m).
\end{eqnarray*}
The functions $F(-\bar m)$ and $\psi(x,y,m)$ are
smooth on the left and right-hand sides of complex
plane of $m$ and one can show that the integral
exists. Let us represent it into more convenient form.
For this we rewrite the integral into sum of two
integrals over left-hand side and right-hand side of
complex plane. Let us integrate by parts these
integrals. As a result we obtain
\begin{eqnarray*}
\pt_k\varphi={-1\over2i\pi}\sum_{\pm}\int_{\pt
\O^{\pm}}{(\pm)d\bar m \over k-m} \big[f(-\bar m)\times\\
\exp(i(m+\bar m)x-(m^2-\bar
m^2)y)\psi(x,y,m)\big]\,-\, \\
{-1\over2i\pi}\sum_{\pm}\int\int_{\O^{\pm}}{dm\wedge
d\bar m\over k-m} \bigg[(ix-2my)(\pm)f(-\bar
m)\psi(x,y,m)+ \\  (\pm)\pt_m f(-\bar m)\psi(x,y,m)-
\varphi(x,y,m)f(-\bar m)f(m)\bigg]\times\\
\times\exp(i(m+\bar m)x-(m^2-\bar m^2)y).
\end{eqnarray*}
Here $\O^{\pm}$ is right-hand side (+)
and  left-hand side (-) of complex plane. We mean the
integral over $\pt\O^{+}$ as sum of integral over
right-hand side and left-hand side of the circle with
center at origin of coordinates and with large radius
(which tends into infinity) and of improper integral
over imaginary line and of an integral over  the
circle with small radius $\varepsilon$ with center at
$k=m$. The direction of the integration over
$\pt\O^{+}$ is determined by standard way.
\par
Consider the sum of the integrals over
$\pt\O^{\pm}$. Each of the integrals over large
half-circles is equal to zero as $R\to\infty$ because
the integrand decreases. The sum of the integrals
over the imaginary axis gives doubled integral over
the  imaginary axis in positive direction. The
integral over small circle equals  zero as $\ve\to0$.
\par
Let us evaluate the derivative $\pt_m f(-\bar m)$ of the integrand
of the double integrals. For  this we note that
$\varphi(x,y,k,0)=\psi(x,y,-\bar k,0)$, hence $f(-\bar m)$ has to be
represented by the  function $\psi(x,y,m,0)$ and then: $$ \pt_m
f(-\bar m)={1\over2\pi}\int\int_{\Real^2}dx\,dy \,
u_0(x,y)\big[-f(m)\varphi(x,y,m,0)+ $$ $$
+(-ix-2my)\psi(x,y,m,0)\big]\exp(-i(m+\bar m)x-(m^2-\bar m^2)y). $$
\par
So we obtain a formula $f_{10}(k')$: $$
f_{10}(k')=f_{10}^{(1)}(k')+\sgn[\Re(-k)]f_{10}^{(2)}(k'),
$$ where
\begin{eqnarray*}
f_{10}^{(1)}(k')=-f(-\bar
k')\psi(x,y,k',0){1\over2\pi i}\int\int_{\Real^2} dx
dy u_0(x,y)\times \\ \bigg[ (ix-2ky)\phi(x,y,k,0)+
\\
\bigg( {-1\over2i\pi}\sum_{\pm}V.P.\int_{-i\infty}^{i\infty}{d\bar
m \over k'-m}(\pm)f(-\bar m)\psi(x,y,m,0)- \\
{-1\over2i\pi}\sum_{\pm}\int\int_{\O^{\pm}}{dm\wedge
d\bar m\over k-m} \bigg[(ix-2my)(\pm)f(-\bar
m)\psi(x,y,m)+ \\
(\pm)\pt_m f(-\bar m)\psi(x,y,m)-
\varphi(x,y,m)f(-\bar m)f(m)\bigg] \times \\
\exp(i(m+\bar m)x-(m^2-\bar m^2)y) \bigg)\times\\
\exp(-i(k'+\bar k')x-(k'^2-\bar k'^2)y) \bigg];
\end{eqnarray*}
$$
f_{10}^{(2)}={-1\over\pi}\int\int_{\Real^2}dxdy
u_0(x,y) f(-\bar k')\psi(x,y,k',0). $$ \par An
expression of
$f_{01}(k')=f_{01}^{(1)}(k')+\sgn[\Re(-k)]f_{01}^{(2)}(k')$
has  the form:
\begin{eqnarray*}
f_{01}^{(1)}(k') &
={1\over2i\pi}\int\int_{\Real^2}dxdy\,u_0(x,y)(-ix+2\bar
k'y)\varphi(x,y,k',0) \times
\\
&
\exp(i(k'+\bar k')x-(k'^2-\bar k'^2)y), \\
f_{01}^{(2)}(k') & ={1\over2i\pi}
\int\int_{\Real^2}dxdy\,u_0(x,y)\psi(x,y,k',0) \times
\\
&\times \exp(2i(k'+\bar k')x-2(k'^2-\bar k'^2)y).
\end{eqnarray*}
\par
The expressions of other coefficients of the
expansion have  to be evaluated by the same way. The
lemma is proved.

\section{Asymptotic solution of $\bar D$-problem}

\par
In this section we construct an asymptotic  solution as $t\to\infty$
of the $\bar D$-problem: \bb
\begin{array}{c}
\pt_{\bar k}\mu=\nu\,F(k)\exp(itS),\\
\pt_k\nu=-\mu\,F(-\bar k)\exp(-itS);\\
\end{array}
\left(\begin{array}{c}
\mu\\
\nu
\end{array}\right)|_{|k|\to\infty}
= \left(\begin{array}{c}
1\\
0
\end{array}\right).
\label{bar d1} \ee
\par
The solution of the  problem (\ref{bar d}) has to be obtained using
the solution of (\ref{bar d1}) and the formula: \bb \left(
\begin{array}{c}
\varphi\\
\psi
\end{array}\right)=
\left(
\begin{array}{c}
\mu(k,\xi,\eta,t)\\
\nu(k,\xi,\eta,t)
\end{array}
\right)+ \left(
\begin{array}{c}
-\nu(-\bar k,\xi,\eta,t)\\ \mu(-\bar k,\xi,\eta,t)
\end{array}
\right). \label{sol bar d} \ee
\par
The asymptotic solution of the problem (\ref{bar d1}) is combined.
Here we construct the asymptotics as $t\to\infty$ uniformly with
respect to all parameters. The stationary points of the functions
$S(k,\bar k,\xi,\eta)$ with respect to parameters $k,\,\,\bar k$
plays the important role in these constructions.
\par
The asymptotic expansion of the solutions is
constructed using the inverse powers of large
parameter $t$ as a asymptotic sequence of asymptotic
expansion outside of small neighborhoods  of the
stationary points of the function. Near the
stationary points the asymptotic  sequence has the
form $t^{-n/2}$, where $n=0,1,2,\dots$. Following the
terminology of matching method \cite{I}, we call the
asymptotics outside of small neighborhoods of
stationary points by outer expansion and we call by
interior expansion the asymptotics near the
stationary points . The right-hand side in the
system  of equations (\ref{bar d1}) is discontinuous
on the  line $\Re(k)=0$, therefore the asymptotic
solution is differentiable out of the line $\Re(k)=0$.
\par
The domains of validity for interior and exterior asymptotic
expansions are intersected. This fact is used by matching method in
order to construct unique combined asymptotic expansions.
\par
The phase function $S$ depends on two parameters
$(\xi,\eta)$ $\in \Real^2$. On the curve
$12\xi+\eta^2=0$ the confluence of two stationary
points of the function $S$ occurs. In this case we
have one confluent stationary point. The structure of
the asymptotic expansion of solution of (\ref{bar
d1}) is changed here. As $|12\xi+\eta^2|\ll1$ the
expansion is constructed on the powers of $t^{-n/3},$
$n=0,1,2,\dots$ as a asymptotic sequence.
\par
The formulas for the uniform asymptotic expansions of the solution
of (\ref{bar d1}) are large and it seems convenient  to  formulate
the results about general case in section 4.1 and about confluence
case in section  4.2 for  convenience  .

\subsection{Asymptotics in a general case}

\par
In this section the combined asymptotic solution of (\ref{bar d1})
is constructed when the phase function $S$ has   nondegenerate
stationary points $k=k_1$ and $k=k_2$. Here we suppose that $\forall
k\in\Complex, \,t^{1/3}\big|\pt^2_kS\vert_{k=k_{1,2}}\big|\gg1$ as
$t\to\infty$. This leads to restriction on values of the parameters
$\xi$ and $\eta$, namely,  $t^{1/3}|12\xi+\eta^2|\gg1$. The
asymptotic expansion which is  uniform with respect to
$k\in\Complex$ is formulated in the end of this section.
\par
To construct the combined asymptotic solution which
is valuable as $k\in\Complex$ we obtain exterior and
interior asymptotic expansions. These expansions are
valid outside of small neighborhoods of $k_j$,
$j=1,2$ and in the small neighborhoods of $k_j$
respectively. Denote: $\theta =
\sqrt{-12\xi-\eta^2}$.  Then one can obtain an
expressions for the stationary points of $S$ by using
the parameter $\theta$:
$k_1={1\over12}(i\eta+\theta),\quad
k_2={1\over12}(i\eta-\theta)$.

\subsubsection{The stationary points outside of the break line}

\par
Consider the case when $\Re(\theta)\not=0$, i.e. when the stationary
points $k_{1,2}$ are outside of the discontinuity line $\Re(k)=0$.
Let us formulate a result of this section about the combined
asymptotic solution:

\begin{lemma}
Let the system of the equations (\ref{bar d1}) have
not the homogeneous solutions, $F(k)\in C^2\cap L_1$
as $\Re(k)\not=0$ and the parameters $\xi$ and $\eta$
satisfy the inequality $t^{1/3}|\theta|^2\gg1$, then:
\newline
${}\qquad{}$ when $\sqrt{t|\theta|}|k-k_{1,2}|\gg1$ the formal
asymptotic solution of the system (\ref{bar d1}) with respect to
$mod(O(t^{-2}|\pt_{k}S|^{-3}))$ has the form: $$
\tilde\mu=1+t^{-1}\stackrel{1}{\mu}(k,\xi,\eta), $$ $$
\tilde\nu=(t^{-1}\stackrel{1}{\nu}_1(k,\xi,\eta)+
t^{-2}\stackrel{2}{\nu}_1(k,\xi,\eta))\exp(-itS)\,+
\,t^{-1}\stackrel{1}{\nu}_0(k,\xi,\eta); $$ the functions
$\stackrel{1}\mu,\,\stackrel{1}{\nu}_1,\,\stackrel{2}{\nu}_1,
\,\stackrel{1}{\nu}_0$ are defined by (\ref{ex-mu1}),
(\ref{ex-nu1}), (\ref{ex-nu2}), (\ref{ex-nu0});
\newline
${}\qquad{}$ when $|\theta|^{-1}\,|k-k_j|\ll1$ the
formal asymptotic solution of the system (\ref{bar
d1}) with respect to $mod(O(t^{-1}|\theta|^{-1}))$
has the form: $$
\tilde\mu=1+t^{-1}\stackrel{1}{M}(l_j,\xi,\eta), $$ $$
\tilde\nu=\big(t^{-1/2}\stackrel{1}{N}(l_j,\xi,\eta)+
t^{-1}\stackrel{2}{N}(l_j,\xi,\eta)\big)\exp(-itS),
$$ where $l_j\,, j=1,2$, are defined by formula: $$
l_j=\sqrt{t}\big(k-k_j\big)\sqrt{{\pt^2_k S_j\over
2}\,+\, 4(k-k_j)}, $$ the functions
$\stackrel{1}{M}(l_j,\xi,\eta),\,
\stackrel{1}{N}(l_j,\xi,\eta),\,
\stackrel{2}{N}(l_j,\xi,\eta)$ are defined by
(\ref{in-M1}), (\ref{in-N1}), (\ref{in-n2}).
\end{lemma}
\par
{\large Proof.} Let us construct the external
asymptotic solution in the form:  \bb
\mu^{ex}=1+t^{-1}\stackrel{1}{\mu}(k,\xi,\eta)+
t^{-2}\stackrel{2}{\mu}_1(k,\xi,\eta)\exp(itS)+\dots,
\label{ex-mu} \ee
\begin{eqnarray}
\nu^{ex}=\big(t^{-1}\stackrel{1}{\nu}_1(k,\xi,\eta)+
t^{-2}\stackrel{2}{\nu}(k,\xi,\eta)+\dots\big)\exp(-itS)+
\nonumber\\
+t^{-1}\stackrel{1}{\nu}_0(k,\xi,\eta)+\dots. \label{ex-nu}
\end{eqnarray}
Let us  substitute the formulas (\ref{ex-mu}) and
(\ref{ex-nu}) into (\ref{bar d1}), equate
coefficients with equal power of $t$. As a result we
obtain:
\begin{eqnarray*}
-i\pt_k S\stackrel{1}{\nu}_1(k,\xi,\eta)=
-\sgn(\Re(k))f(-\bar k), \qquad\qquad\qquad\qquad\\
\pt_{\bar k}\stackrel{1}{\mu}(k,\xi,\eta)+ i\pt_{\bar
k} S\stackrel{2}{\mu}_1(k,\xi,\eta)\exp(itS) =
\qquad\qquad\qquad\quad\\
-\sgn(\Re(k))f(k)\stackrel{1}{\nu}_1(k,\xi,\eta)+ \\
\sgn(\Re(k))f(k)\stackrel{1}{\nu}_0(k,\xi,\eta)\exp(itS),
\\
 -i\pt_k S\exp(-itS)\stackrel{2}{\nu}_1(k,\xi,\eta)+
\pt_k\stackrel{1}{\nu}_0(k,\xi,\eta)= \qquad\qquad\qquad\\
\big(-\sgn(\Re(k))f(-\bar
k)\stackrel{1}{\mu}(k,\xi,\eta)
\pt_k\stackrel{1}{\nu}_1(k,\xi,\eta)\big)\exp(-itS).
\end{eqnarray*}
If we equate to zero the coefficients of oscillated
terms and nonoscillated terms  of the expansions
(\ref{ex-mu}) and  (\ref{ex-nu}) respectively , then
we obtain formulas: \bb \stackrel
{1}{\nu}_1={\sgn(\Re(k))f(-\bar k)\over i\pt_k S};
\label{ex-nu1} \ee \bb
\pt_k\stackrel{1}{\nu}_0(k,\xi,\eta)=0;
\label{ex-nu10} \ee \bb \pt_{\bar k}\stackrel
{1}{\mu}= {-f(-\bar k)f(k)\over i\pt_k S};
\label{ex-mu1equat} \ee \bb
\stackrel{2}{\mu}_1(k,\xi,\eta)=
{\sgn(\Re(k))f(k)\stackrel{1}{\nu}_0(k,\xi,\eta)
\over i\pt_{\bar k}S}; \label{ex-mu20} \ee \bb
\stackrel{2}{\nu}={-1\over i\pt_k S}
\left(-\sgn(\Re(k))f(-\bar k)\stackrel
{1}{\mu}\pt_k\stackrel{1}{\nu}_1\right).
\label{ex-nu2} \ee
\par
The function $\stackrel{1}{\nu}_1$ has a jump on the
imaginary axis of $k$. This jump is ${2f(k)\over
i\pt_k S}$. In order for the coefficients of the
asymptotics of function $\nu$ as $t^-1$ to be
continuous, we add an analytic function of $\bar k$,
which has the same jump on the imaginary axis in
inverse direction, to $\stackrel{1}{\nu}_1\exp(iSt)$:
\bb \stackrel{1}{\nu'}_0(\bar k) ={1\over
i\pi}\int_{-i\infty}^{i\infty} {dn f(n)\over (\bar k
-n)\pt_n S}. \label{ex-nu1'} \ee
\par
The function $\stackrel{1}{\nu'}\!\!_1$ define  an
analytic function $\stackrel{1}{\nu}_0$  of $\bar k$
uncompletely. The rest terms will be defined when we
will match  the exterior and interior expansion.
\par
A Cauchy-Green formula  gives solution of
(\ref{ex-mu1equat}) which is decreasing as
$|k|\to\infty$ and bounded when $\forall k\in
\Complex$: \bb \stackrel {1}{\mu}={-1\over 2i\pi}
\int\int_{\Complex}{ dp\wedge d\bar p  \over k-p}
{f(-\bar p)f(p)\over i\pt_p S}. \label{ex-mu1} \ee
\par
We can obtain  the domain of values of $k$, where the
external expansion is valid, using the condition
$|t^{-1}\stackrel {1}{\nu}_1/(t^{-2}\stackrel
{2}{\nu})|\gg1$. As a result of calculations we
obtain: $$ \sqrt{t|\theta|}|k-k_j|\gg1. $$
\par
Let us construct the interior asymptotic solution
which is valid in the neighborhood of point $k_j$ as
$j=1,2$. Denote new scaling variable by $l_j$: \bb
l_j^2=t(k-k_j)^2{\pt_k^2S_j\over 2}\,+\,4t(k-k_j)^3.
\label{lj} \ee
\par
When $|l_j|$ is not so large ($t^{1/2}|\pt_k^2
S_{j}|^{3/2}\gg|l_j|$) an asymptotic formula is valid
as $t\to\infty$ \bb (k-k_j)=\sqrt{2\over t\pt_k^2S_j}
l_j\,-\,{8\over t(\pt_k^2S_j)^2} l_j^2+\dots.
\label{klj} \ee
\par
Rewrite the system (\ref{bar d1}) into terms of new variables $l_j$
and  $\bar l_j$. Substitute the asymptotic expansion \bb
\mu^{in}=1+t^{-1}\stackrel{1}{M}_j(l_j,\xi,\eta)+\dots,
\label{mu-in1} \ee \bb
\nu^{in}=(t^{-1/2}\stackrel{1}{N}_j(l_j,\xi,\eta)+
t^{-1}\stackrel{2}{N}_j(l_j,\xi,\eta)+\dots)\exp(-itS),
\label{nu-in1} \ee into system (\ref{bar d1}).
\par
As a result one can obtain equations for the coefficients of
expansions (\ref{mu-in1})  and (\ref{nu-in1}): \bb
\pt_{l_j}\stackrel{1}{N}_{j}-2il_j\stackrel{1}{N}_{j}=
-\sqrt{2\over\pt^2_k S_j} \sgn(\Re(k_j))f(-\bar k_j);
\label{N1equat} \ee $$
\pt_{l_j}\stackrel{2}{N}_{j}-2il_j\stackrel{2}{N}_{j}=
\sgn(\Re(k_j))\Bigg[\Big({16\over(\pt^2_k S_j)^2}\, f(-\bar k_j)-
{2\over\pt^2_k S_j}\,f_{10}(-\bar k_j)\Big)l_j- $$ \bb - {2\bar
l_j\over|\pt^2_k S_j|}f_{01}(-\bar k_j)\Bigg]; \label{N2equat} \ee
\bb \pt_{\bar
l_j}\stackrel{1}{M}_{j}=-\sgn(\Re(k_j))f(k_j)\stackrel{1}{N}_{j}
\sqrt{{2\over\pt^2_{\bar k} S_j}}. \label{M1equat} \ee
\par
Construct the solutions of the equations
(\ref{N1equat})--(\ref{M1equat}). Since the external
expansion has not the terms of order $t^{-1/2}$, then
the boundary condition for function
$\stackrel{1}{N}_1(l_j,\xi,\eta)$ has the form: $$
\stackrel{1}{N}_j(l_j,\xi,\eta)|_{|l_j|\to\infty}=0.
$$ The solution of the boundary condition for
$\stackrel{1}{N}_j(l_j,x,y)$ is evaluated by formula:
\begin{eqnarray}
\stackrel{1}{N}_j(l_j,\xi,\eta)
=\sgn(\Re(k_j)) {\sqrt{2}f(-\bar k_j)\over\sqrt{\pt^2_k S_j}}
{\exp(i(l_j^2+\bar l_j^2))\over2i\pi}\times \nonumber\\
\times \int\int_{\Complex} {dn\wedge d\bar n\over
\overline{l_j-n}}\exp(-i(n^2+\bar n^2)); \label{in-N1}
\end{eqnarray}
\par
The solution of nonhomogeneous  equation for the function
$\stackrel{2}{N}_j(l_j,\xi,\eta)$ has the form:
\begin{eqnarray}
\stackrel{2}{N}_j\!\!\!^s(l_j,\xi,\eta) = \sgn(\Re(k_j))\bigg[
{8f(-\bar k_j)\over (\pt_k^2S_j)^2}\,-\,
{f_{10}(-\bar k_j)\over\pt_k^2 S_j}\nonumber\\
{2f_{01}(-\bar k_j)\over |\pt_k^2 S_j|}\exp(i(l_j^2+\bar l_j^2))
\nonumber\\
\,-\, \bar l_j{2 f_{01}(-\bar k_j)\over|\pt_k^2 S_j|}
{\exp(i(l_j^2+\bar l_j^2))\over2i\pi} \int\int_{\Complex}{dn\wedge
d\bar n\over \overline{l_j-n}}\exp(-i(n^2+\bar n^2)) \bigg].
\label{in-n2s}
\end{eqnarray}
\par
The partial solution of the equation for the function
$\stackrel{1}{M}_j(l_j,\xi,\eta)$ has to be written
as four-multiple integral: \bb
\stackrel{1}{M}_{j}\!\!^s= {-2f(k_j)f(-\bar k_j)\over
|\pt^2_{\bar k} S_j|}J, \label{in-M1s} \ee where
\begin{eqnarray*}
J={1\over 2i\pi}\int\int_{\Complex}{dn\wedge d\bar
n\over l_j-n}\,
{\exp(i(n^2+\bar n^2))\over2i\pi}\times\qquad\qquad\qquad\\
\int\int_{\Complex} {dm\wedge d\bar m\over
\overline{n-m}}\exp(-i(m^2+\bar m^2)).
\end{eqnarray*}
\par
It is possible to reduce this four-multiple integral
into two-multiple integral (see Appendix):
\begin{eqnarray*}
J=\bar l_j{\exp(i(l_j^2+\bar l_j^2))\over 2i\pi}
\int\int_{\Complex} {dn\wedge d\bar n\over
\overline{l_j-n}}\exp(-i(n^2+\bar n^2))\,-\,\\
\qquad\qquad\exp(i(l_j^2+\bar l_j^2)).
\end{eqnarray*}
\par
There exists an domain of values of complex parameter
$k$, in which the external and internal asymptotic
expansions  of solution for the problem (\ref{bar
d1}) are valid. In this domain these asymptotic
expansions are equal up to the terms of order
$o(t^{-1})$. In the domain, where
$t^{1/2}|k-k_j|\gg1$ and $|k-k_j|\le t^{-1/4}$, the
external and internal expansion are valid. We compute
asymptotics of external expansion as $k\to k_j$ and
internal expansion as $|l_j|\to\infty$.
\par
Let us present the asymptotics of the functions
$\stackrel{1}{N}_{j}$ and
$\stackrel{2}{N}_{j}\!\!^{s}$ as $|l_j|\to\infty$:
\begin{eqnarray*}
\stackrel{1}{N}_j|_{|l_j|\to\infty}=\,-\,\sgn(\Re(k_j))
\sqrt{2\over\pt_k^2 S_j}f(-\bar
k_j)\times\qquad\qquad\qquad\\
\left({1\over2il_j}\,+\,{\exp(i(l_j^2+\bar
l_j^2))\over\bar l_j}\,+\,O(|l_j|^{-3})\right).
\\
\stackrel{2}{N}_j(l_j,\xi,\eta)|_{|l_j|\to\infty}=
\sgn(\Re(k_j))\left[ \left( {8if(-\bar k_j)\over(\pt^2_k S_j)^2}\,-\,
{if_{10}(-\bar k_j)\over\pt^2_k S_j} \right)\right.
\\
\left. + {\bar l_j\over 2il_j} {-2 f_{01}(-\bar
k_j)\over|\pt^2_{\bar k} S_j|} \,+\,O(|l_j|^{-1}) \right].
\end{eqnarray*}
\par
The matching condition for $\tilde \nu$ means that in
the domain $t^{1/3}|k-k_j|\gg1$ and  $|k-k_j|=o(1)$:
$$ \left(t^{-1/2}\stackrel{1}{N}_j(l_j,\xi,\eta)+
t^{-1}\stackrel{2}{N}_j(l_j,\xi,\eta)\right)\exp(-itS)\,-\,
$$ $$
t^{-1}\left(\stackrel{1}{\nu}_1(k,\xi,\eta)\exp(-itS)
+
\stackrel{1}{\nu}_0(k,\xi,\eta)\right)\,=\,o(t^{-1}).
$$ Using the asymptotics of interior expansions of
$\stackrel{2}{N}_j$, which are rewritten in the terms
of external variable $k$ as $k\to k_{1,2}$, and
external expansion of $\stackrel{1}{\nu}_0$, we can
obtain:
\begin{eqnarray} \stackrel{1}\nu_0(k,x,y) &=
\stackrel{1}{\nu'}\!\!_0(\bar k)\,+\, {2f(-\bar
k_1)\exp(itS_1)\over|\pt_k^2 S_1|\overline{(k-k_1)}} +\nonumber\\ &
\,+\, {-2f(-\bar k_2)\exp(itS_2)\over|\pt_k^2
S_2|\overline{(k-k_2)}}, \label{ex-nu0}
\end{eqnarray}
where the function $\stackrel{1}{\nu}_0\!\!'(\bar k)$ is defined by
(\ref{ex-nu1'});
\begin{eqnarray}
\stackrel{2}{N}_j= \stackrel{2}{N}_j\!\!^s\,+\,
\stackrel{1}{\nu'}\!\!_0(k_j)\exp(i(l_j^2+\bar l_j^2))\exp(itS_j)\,
+\,\nonumber\\ +\,{2\sgn(\Re(k_m))f(-\bar
k_m)\exp(itS_m)\over|\pt_k^2 S_m|\overline{(k_j-k_m)}}
\exp(i(l_j^2+\bar l_j^2)), \label{in-n2}
\end{eqnarray}
where  $m\not=j$, the function $\stackrel{2}{N}_j\!\!^s$ is defined
by (\ref{in-n2s}).
\par
Let us  construct $\stackrel{1}{M}_j(l_j,\xi,\eta)$.
The asymptotics of (\ref{in-M1s}) as $|l_j|\to\infty$
has the form: $$
\stackrel{1}{M}_{j}\!\!^s|_{|l_j|\to\infty}= {\bar
l_j\over l_j}\sqrt{{\pt_k^2 S_j\over\pt^2_{\bar k}
S_j}}{i\over 12\pi} {f(-\bar k_j) f(k_j)\over\pt^2_k
S}. $$
\par
Evaluate the asymptotics of $\stackrel{1}\mu_0$ as $k\to k_j$: $$
\stackrel{1}\mu_0|_{k\to k_j}={\overline{(k-k_j)}\over k-k_j}
{f(-\bar k_j) f(k_j)\over12i(k_j-k_n)}\,-\,{f(k_j)f(-\bar k_j)
\over12i(k_j-k_n)}\,+ $$ $$ +\,{1\over
2i\pi}\int\int_{\Complex}dp\wedge d\bar p {f(-\bar p)
f(p)-f(k_j)f(-\bar k_j)\over 12i(p-k_j)^2(p-k_n)}, $$ here
$n\in{1,2},\, n\not=j$.
\par
The matching condition for $\tilde \mu$ when $t^{-1/3}|k-k_j|\gg1$
and $|k-k_j|=o(t^{-1/4})$ has the form: $$
(1+t^{-1}\stackrel{1}\mu_0(k,\xi,\eta))
(1+t^{-1}\stackrel{1}{M}_j(l_j,\xi,\eta))=o(t^{-1}).
$$
\par
These condition allows to define
$\stackrel{1}{M}_j(l_j,\xi,\eta)$: \bb
\stackrel{1}{M}_j(l_j,\xi,\eta)=
\stackrel{1}{M}_j\!\!^s(l_j,\xi,\eta)+C_j(\xi,\eta).
\label{in-M1} \ee Here the function
$\stackrel{1}{M}_j\!\!^s(l_j,\xi,\eta)$ is defined by
(\ref{in-M1s}), $C_j(\xi,\eta)$ has the form:
\begin{eqnarray} C_j(\xi,\eta)=-\,{f(k_j)f(-\bar k_j)
\over12i(k_j-k_n)}\,+\,\qquad\qquad\qquad\nonumber\\{1\over
2i\pi}\int\int_{\Complex}dp\wedge d\bar p {f(-\bar p)
f(p)-f(k_j)f(-\bar k_j)\over 12i(p-k_j)^2(p-k_n)},
\label{Cj}
\end{eqnarray}
where $n\in{1,2},\,
n\not=j$.
\par
Thus we have  constructed the interior expansion in neighborhood of
nondegenerate stationary point of $S$ as $|12\xi+\eta^2|\ge
t^{-1/4}$.
\par
Lemma is proved.
\par
Constructed asymptotic solutions are nonuniform with
respect to $k$. But one can obtain an uniform
asymptotic solution by using their combination. This
uniform solution has the form (see, for example,
\cite{I}): \begin{eqnarray} \left(
\begin{array}{c}
\hat \mu\\ \hat\nu
\end{array}
\right)= \left(
\begin{array}{c}
\tilde \mu\\ \tilde\nu
\end{array}
\right) \,+\, \left(
\begin{array}{c}
\tilde M_1\\ \tilde N_1
\end{array}
\right) \,+\, \left(
\begin{array}{c}
\tilde M_2\\ \tilde N_2
\end{array}
\right) \,-\,\qquad\qquad\nonumber\\
\qquad\qquad A_{1,k} \left(
\begin{array}{c}
\tilde M_1\\ \tilde N_1
\end{array}
\right)\,-\, A_{1,k} \left(
\begin{array}{c}
\tilde M_2\\ \tilde N_2
\end{array}
\right). \label{matching1} \end{eqnarray}
\par
The result of action of operator $A_{n,k}$ on the function $\tilde
{M}$ is defined as follows (\cite{I}). Take the formula for $\tilde
{M}(l_j,\xi,\eta,t)$, and change the dependence on variable $l_j$
into the dependence on variable $k$ using the formula (\ref{lj}).
Rewrite the sum of all terms of asymptotic expansion up to $t$ with
powers equal to  ${-m}$, where $0\le m\le n$. For example for the
functions $\tilde M(l_1, \xi,\eta,t)$ and $\tilde
N(l_1,\xi,\eta,t)$ this process leads to the formulas: $$
A_{1,k}(\tilde
M_1)=1+t^{-1}\,{\overline{\sqrt{\theta(k-k_1)^2+4(k-k_1)^3}} \over
\sqrt{\theta(k-k_1)^2+4(k-k_1)^3}}\, {f(-\bar k_1)f(k_1)\over
2|\pt_k^2 S_1|}; $$
\begin{eqnarray*}
A_{1,k}(\tilde N_1) =t^{-1} \Bigg[
{f(-\bar k_1)\over 2i\theta\sqrt{\theta(k-k_1)^2+4(k-k_1)^3}} -\\
\,-\, {f(-\bar k_1)\exp(-it(S-S_1)) \over
|\theta|\overline{\sqrt{\theta(k-k_1)^2+4(k-k_1)^3}}} \,+\,
\\
+\, \stackrel{1}{\nu'}\!\!\!_0 \,-\, {f_{01}(-\bar k_1)\over
|\theta|}\, {\overline{\sqrt{\theta(k-k_1)^2+4(k-k_1)^3}} \over
\sqrt{\theta(k-k_1)^2+4(k-k_1)^3}}\, \Bigg].
\end{eqnarray*}
\par
If we substitute (\ref{matching1}) into (\ref{bar d1}) and evaluate
a remainder, then we obtain:
\begin{theorem}
The formal asymptotic solution of the problem (\ref{bar d1}) with
respect to  $mod\big(O\big((t|\theta|)^{-1}\big)\big)$, which is
uniformly valuable when  $k\in\Complex$ and  $\theta^2 t\gg1$, has
the form (\ref{matching1}).
\end{theorem}

\subsubsection{The stationary points on the imaginary axis}

\par
If $\Re(\theta)=0$, then the stationary point of the
phase function $S$ belongs to the line, where the
coefficients of the equation (\ref{bar d1}) are
discontinuous. In this case constructing of the
formal asymptotic solution of  problem (\ref{bar d1})
differs from the asymptotic solution which was
constructed before. The main result about the
combined asymptotic solution is:
\begin{lemma}
Let the system of equations (\ref{bar d1}) have no homogeneous
solutions, $F(k)\in C^2\cap L_1$ as $\Re(k)\not=0$; the parameters
$\xi$ and $\eta$ are $-t^{2/3}(12\xi+\eta^2)\gg1$, then: \newline
$\quad\qquad$ when $\sqrt{t|\theta|}|k-k_{1,2}|\gg1$the formal
asymptotic solution of (\ref{bar d1}) with respect to
$mod(O(t^{-2}|\pt_{k}S|^{-3}))$ has the form: $$
\tilde\mu=1+t^{-1}\stackrel{1}{\a}(k,\xi,\eta), $$ $$
\tilde\nu=(t^{-1}\stackrel{1}{\beta}_1(k,\xi,\eta)+
t^{-2}\stackrel{2}{\beta}_1(k,\xi,\eta))\exp(-itS)\,+
\,t^{-1}\stackrel{1}{\beta}_0(k,\xi,\eta); $$ the functions
$\stackrel{1}{\a},\,\stackrel{1}{\b}_1,\,\stackrel{2}{\b}_1,
\,\stackrel{1}{\b}_0$ are defined by (\ref{ex-a1}), (\ref{ex-b1}),
(\ref{ex-b2}), (\ref{ex-b01});
\newline $\quad$ when $|\theta|\,|k-k_j|\ll1$ the
formal asymptotic solution of (\ref{bar d1}) with
respect to $mod(O(t^{-1}|\theta|^{-1}))$ has the
form: $$ \tilde
Y=1+t^{-1}\stackrel{1}{Y}(l_j,\xi,\eta), $$ $$ \tilde
Z=\big(t^{-1/2}\stackrel{1}{Z}(l_j,\xi,\eta)+
t^{-1}\stackrel{2}{Z}(l_j,\xi,\eta)\big)\exp(-itS),
$$ where the functions
$\stackrel{1}{Y}(l_j,\xi,\eta),\,
\stackrel{1}{Z}(l_j,\xi,\eta),
\,\stackrel{2}{Z}(l_j,\xi,\eta)$ are defined by
(\ref{in-Y1}), (\ref{in-Z12}).
\end{lemma}
\par
{\bf The proof.} Let us construct the asymptotics.
The external expansion is constructed  similarly as
in 4.1.1. The main difference is that the  stationary
points of the function $S$ are on the line of
discontinuity of the coefficients of system (\ref{bar
d1}). It leads to sufficient modifications in the
formulas for asymptotics. Let us  find the external
expansion in the form: \bb
\mu^{ex}=1+t^{-1}\stackrel{1}{\a}(k,\xi,\eta)+
t^{-2}\stackrel{2}{\a}_1(t,\xi,\eta)\exp(itS)+\dots,
\label{ex-a} \ee \bb
\nu^{ex}=t^{-1}(\stackrel{1}{\b}(k,\xi,\eta)\exp(-itS)+
\stackrel{1}{\b}_0(k,\xi,\eta))+t^{-2}\stackrel{2}{\b}\exp(-itS)+\dots.
\label{ex-b} \ee After substituting of (\ref{ex-a})
and (\ref{ex-b}) into the system (\ref{bar d1}) we
equate the coefficients with identical powers of $t$
and with oscillated  and nonoscillated terms
correspondingly. As a result we  obtain equations for
the coefficients  of asymptotic expansions
(\ref{ex-a}) and (\ref{ex-b}). The obvious formulas
are \bb
\stackrel{1}{\a}={-1\over2i\pi}\int\int_{\Complex}{dn\wedge
d\bar n\over k-n}{f(-\bar n)f(n)\over i\pt_n
S(n,\xi,\eta)}, \label{ex-a1} \ee \bb
\stackrel{1}{\b}_1={\sgn(\Re(k))f(-\bar k)\over
i\pt_kS}. \label{ex-b1} \ee \par The function
$\stackrel{1}{\b}_1$ is discontinuous on the
imaginary axis. We add an analytic function with the
same jump on the  imaginary axis in back direction
then the coefficient of the expansion (\ref{ex-b}) at
$t^{-1}$ is continuous: \bb
\stackrel{1}{\b'}_0={1\over\pi i\pt_{\bar
k}S}\int_{-i\infty}^{i\infty} {dn f(n)\pt_{\bar n}
S\over (\bar k-n)\pt_n S} = {1\over \pi i \pt_{\bar
k}S}\int_{-i\infty}^{i\infty} {dn f(n)\over\bar k-n}.
\label{b1'} \ee Unlike $\stackrel{1}{\nu'}_0$, this
function have singularities on  the discontinuous
line.
\par
The formulas for the other coefficients of the expansions
(\ref{ex-a}) and (\ref{ex-b}) have the forms: \bb
\stackrel{2}{\a}_1={\sgn(\Re(k))f(k)\stackrel{1}{\b}_0(k,\xi,\eta)
\over i\pt_{\bar k}S}, \label{ex-a20} \ee \bb
\stackrel{2}{\b}={-1\over i\pt_k S}\bigg(-\sgn(\Re(k)f(-\bar
k)\stackrel{1}{\a}- \pt_{k}\stackrel{1}{\b}_1\bigg). \label{ex-b2}
\ee Here the analytic function $\stackrel{1}{\b}_0$ is  still
undefined. We will finally  define this function  after  matching of
external  and  internal expansions.
\par
The interior expansion in the neighborhood of the
point $k_j$ depend on the scaling variable $l_j$. The
asymptotics of the functions $\mu$ and $\nu$ has the
same asymptotic sequence, but the equations for
$\stackrel{1}{Z}_j$, $\stackrel{2}{Z}_j$ and
$\stackrel{1}{Y}_j$ have discontinuous right-hand
sides.
\begin{eqnarray}
\pt_{l_j}\stackrel{1}{Z}_{j}-2il_j\stackrel{1}{Z}_{j}=
\sqrt{2\over\pt^2_k S_j} f(-\bar k_j)\sgn(\Re(\bar l\exp(i\pi/4)));
\label{Z1equat}
\\
\pt_{l_j}\stackrel{2}{Z}_{j}-2il_j\stackrel{2}{Z}_{j}=
\sgn(\Re[\bar l\exp(i\pi/4)])\times{}\qquad\qquad\qquad\qquad\qquad{}
\nonumber\\
\times \Big({16\over(\pt^2_k S_j)^2}\,  f_{00}(-\bar k_j)+
{2l_j\over\pt^2_k S_j}\,f_{01}^{(1)}(-\bar k_j) +
{2i\bar l_j\over|\pt^2_k S_j|}f_{10}^{(1)}(-\bar k_j)\Big)+ \nonumber\\
+ {2l_j\over\pt_k^2 S_j}f^{(2)}_{01}(-\bar k_j)+{2i\bar
l_j\over|\pt_k^2S_j|} f^{(2)}_{10}(-\bar k_j). \label{Z2equat}
\\
\pt_{\bar l_j}\stackrel{1}{Y}_{j}=\sgn(\Re(l\exp(-i\pi/4)))
f(k_j)\stackrel{1}{Z}_{j}\sqrt{{2\over\pt^2_{\bar k} S_j}}.
\label{Y1equat}
\end{eqnarray}
\par
Continuous partial solutions of the equations
(\ref{Z1equat}), (\ref{Z2equat}) and (\ref{Y1equat})
are obtained by Cauchy-Green formula. The formulas
for the partial solutions of the equations
(\ref{Z1equat}) and (\ref{Z2equat}) are differing from
the formulas obtained in section 4.1.1 by function
$\sgn(-\Re(\bar n\exp(i\pi)))$ under the sign of
double integral: \bb \stackrel{1}{Z}_j\!\!^s={\cal
J}[\stackrel{1}{H_j}](l_j),\quad
\stackrel{2}{Z}_j\!\!^s={\cal
J}[\stackrel{2}{H_j}](l_j), \label{Z12} \ee where
$\stackrel{1}{H}_j$ and $\stackrel{2}{H}_j$ are the
right-hand side of the equations (\ref{Z1equat}) and
(\ref{Z2equat}). An operator ${\cal J}$ has the form:
$$ {\cal J}[h](l_j)={\exp(i(l_j^2+\bar
l_j^2))\over2i\pi} \int\int_{\Complex}{dn\wedge d\bar
n\over \overline{l_j-n}} h(n)\exp(-i(l_j^2+\bar
l_j^2)). $$
\par
The partial solution of (\ref{Y1equat}) contains of
the double integral the same as in (\ref{in-M1s}) and
the term, which is defined by integral over the
discontinuous line. The function
$\stackrel{1}{Y}_j\!\!\!^s$ has the form: \bb
\stackrel{1}{Y}_j\!\!\!^s={2f(k_j)f(-\bar k_j)\over
|\pt_{\bar k}^2 S_j|} \big(J-2J_{+-}-2J_{-+}\big),
\label{Y1s} \ee where:
\begin{eqnarray*}
J_{\mp\pm}=\int\int_{\O^\mp}{dm\wedge d\bar m\over
l-m} {\exp(i(m^2+\bar m^2))\over
2i\pi}\times\qquad\qquad\\
\int\int_{\O^\pm}{dn\wedge d\bar n\over
\overline{n-m}}\exp(-i(n^2+\bar n^2)).
\end{eqnarray*}
Here $\O^{\pm}=\{\pm\Re[l\exp(-i\pi/4)>0\}$. Note that
$J_{-+}(l)=J_{+-}(-l)$. In the Appendix we show,
that  we can reduce the integral  $J_{-+} $ into
double integral. As a result we obtain:
\begin{eqnarray*}
J_{-+}=-{3\over 4}i\pi + \qquad\qquad\qquad
\qquad\qquad\qquad\qquad\qquad\qquad\qquad\\
il\int\int_{\O^+} {dm\wedge
d\bar m\over il-\bar m}
\exp(-i(m^2+\bar m^2)), \,\, {\hbox{when}}\,\, l\in\O^+; \\
J_{-+}=\bar l\exp(i(l^2+\bar l^2))\int\int_{\O^+}
{dm\wedge d\bar m\over il-\bar m} \exp(-i(m^2+\bar
m^2))\qquad
\\
-{1\over 4}i\pi - {1\over 2}i\pi \exp(i(l^2+\bar l^2)), \quad
{\hbox{when}}\quad l\in\O^-;
\end{eqnarray*}
\par
Thus, the function $\stackrel{1}{Y}_j\!\!\!^s$ is represented by sum
of double integrals.
\par
Let us consider matching of the external and internal
asymptotic expansions. For that we need the
asymptotic behaviour of the partial solutions of
(\ref{Z1equat}) and (\ref{Z2equat}) as $|l|\to\infty$.
The asymptotic behaviour of the double integrals is
evaluated in the Appendix. By using it we obtain:
\begin{eqnarray*}
\stackrel{1}{Z}_j\!\!^s\vert_{|l|\to\infty}= \sgn\big[\Re[-\bar
l_j\exp(i\pi/4)]\big] f(-\bar k_j)\sqrt{2\over\pt^2_k S_j}
\bigg({-1\over 2il_j}\,+\, \\
\,+\, {\exp(-i(l^2_j+\bar l^2_j))\over2\bar l_j}\,+O(|l_j|^{-2})
\bigg);
\\
\stackrel{2}{Z}_j\!\!^s\vert_{|l_j|\to\infty}=
\sgn\big[\Re[-\bar l_j\exp(i\pi/4)]\big]
\times\qquad\qquad\qquad\qquad\\
\bigg[\Bigg({-\bar
l_j\over2il_j}-{1\over2}\exp(i(l_j^2+\bar
l_j^2))\bigg) {2if_{10}^{(1)}\over|\pt^2_kS_j|^2}+
\\
\,+\, \bigg({1\over 4}\,-\,
{1\over4}\exp(i(l_j^2+\bar l_j^2))\bigg)
2f_{01}^{(1)} \Bigg] +\qquad\qquad\\
{2if_{10}^{(2)}\over|\pt_k^2S_j|} \bigg({\bar
l_j\over -2il}+ {1\over2}\exp(i(l^2_j+\bar
l^2_j))\bigg) + if_{01}^{(2)}+ O(|l_j|^{-1}).
\end{eqnarray*}
The asymptotic behaviour  of the function
$\stackrel{1}{\b}_0\!\!\!''$ is:
\begin{eqnarray*}
\stackrel{1}{\b}_0\!\!\!''|_{k\to k_j}=
{1\over12i\overline{(k-k_j)(k_j-k_m)}} \bigg[
f(k_j)\sgn[\Re(-\bar k)]+\qquad\qquad
\\
+{1\over\pi i}\int_{-\infty}^{\infty}dn
{f(in)-f(k_j)\over n-\overline{ik_j}} \bigg]+
\qquad\\
{-1\over12i(k_j-k_m)^2} \bigg[
f(k_j)\sgn[\Re(-\bar k)]+ \qquad\\
{1\over\pi i}\int_{-\infty}^{\infty}dn
{f(in)-f(k_j)\over n-\overline{ik_j}} \bigg] +
{1\over12i\overline{(k_j-k_m)}} \bigg[
f_{01}^{(1)}(k_j)+
\\+
f_{01}^{(2)}(k_j)\sgn[\Im(\overline{ik})]+
f_{10}^{(1)}(k_j)+f_{10}^{(2)}(k_j)\sgn[\Im(\overline{ik})]\,+\,
\\
+ {1\over\pi i}\int_{-\infty}^{\infty}dn
{f(in)-f(k_j)\overline{(in-ik_j)}
f_{01}^{(2)}(k_j)-(in-ik_j)f_{10}^{(2)}(k_j)
\over (n-\overline{ik_j})^2} \bigg] \\
+o(1).
\end{eqnarray*}
\par
Using the matching conditions for the asymptotics of
the external  $\nu^{ex}$ and internal $\nu^{in}$
expansions we obtain: \bb
\stackrel{1}{\b}_0=\stackrel{1}{\b'}_0 -
{C^1_1\over12i \overline{(k-k_1)(k_1-k_2)}}
\,-\,{C^1_2\over 12i\overline{(k-k_2)(k_2-k_1)}},
\label{ex-b01} \ee where $\stackrel{1}{\b}_0\!\!\!'$
is defined by formula  (\ref{b1'}), $$
C^1_j={1\over\pi
i}\int_{-\infty}^{\infty}dn{f(-n)-f(-k_j)\over
n-\overline{ik_j}},\quad j=1,2; $$ \bb
\stackrel{1}{Z}_j=\stackrel{1}{Z}_j\!\!^s \qquad
\stackrel{2}{Z}_j=\stackrel{2}{Z}\!\!^s_j\,-
\exp(-i(l_j^2+\bar l_j^2))C^2_j, \label{in-Z12} \ee
where functions $\stackrel{1}{Z}_j\!\!\!^s$ and
$\stackrel{2}{Z}_j\!\!\!^s$ are defined by
(\ref{Z12}),
\begin{eqnarray*}
C^1_j={i\over\pt_k^2S_j}f_{10}^{(2)}(-\bar k_j)\,-\,
\qquad\qquad\qquad \\
{1\over\pi i}\int_{-\infty}^{\infty}dn{f(n)-f(-\bar
k_j)-f_{01}^{(2)}(-\bar
k_j)\overline{(n-ik_j)}(n-ik_j)f_{10}^{(2)}(k_j)\over
(n-\overline{ik_j})^2}.
\end{eqnarray*}
\par
The asymptotic behaviour of the function $\stackrel{1}{Y}_j\!\!\!^s$
is : $$ \stackrel{1}{Y}_j\!\!\!^s=\bigg(-i\pi+{i\bar l_j\over12\pi
l_j}\bigg) {f(-\bar k_j)f(k_j)\over|\pt_k^2S_j|}. $$ The matching
condition for the external expansion of $\mu^{ex}$ and for the
internal expansion of $\mu^{in}$ gives: \bb
\stackrel{1}{Y}_j=\stackrel{1}{Y}\!\!^s_j\,+\,C_j(\xi,\eta)+
i\pi{f(-\bar k_j)f(k_j)\over|\pt_k^2S_j|}, \label{in-Y1} \ee where
$\stackrel{1}{Y}_j\!\!^s$ is defined by (\ref{Y1s}), The function
$C_j(\xi,\eta)$ is defined by (\ref{Cj}).
\par
The lemma is proved.
\par
Constructed  external and internal expansions are irregular with
respect to parameters $k,\xi,\eta$. The uniform expansion  with
respect to $k$ when$(12\xi+\eta^2)t^{1/3}\gg1$ has to be constructed
by the same way as in preceding section: \bb \left(
\begin{array}{c}
\hat \mu\\ \hat\nu
\end{array}
\right)= \left(
\begin{array}{c}
\tilde \mu\\ \tilde\nu
\end{array}
\right) \,+\, \left(
\begin{array}{c}
\tilde Y_1\\ \tilde Z_1
\end{array}
\right) \,+\, \left(
\begin{array}{c}
\tilde Y_2\\ \tilde Z_2
\end{array}
\right) \,-\, A_{1,k} \left(
\begin{array}{c}
\tilde Y_1\\ \tilde Z_1
\end{array}
\right)\,-\, A_{1,k} \left(
\begin{array}{c}
\tilde Y_2\\ \tilde Z_2
\end{array}
\right). \label{matching2} \ee The operator $A_{n,k}$ was defined
above. Let us substitute (\ref{matching2}) into (\ref{bar d1}) and
evaluate a remainder. As a result we obtain

\begin{theorem}
The formal asymptotic solution of the problem
(\ref{bar d1}) with respect to
$mod\big(O\big((t|\theta|)^{-1}\big)\big)$, which is
uniform when $k\in\Complex$ and \linebreak $-\theta^2
t^{2/3}\gg1$, has the form (\ref{matching2}).
\end{theorem}

\subsection{Asymptotics in neighborhood  of confluent stationary point}

\par
The system of equations in the problem (\ref{bar d1})
depends on two control parameters $\xi$ and $\eta$. On
the parabola $12\xi+\eta^2=0$ the degeneracy of
stationary points occurs: $k_1=k_2=k_0=i\eta$. For
this reason the asymptotics constructed in section
4.1 is invalid when parameter
$\theta=\sqrt{\eta^2+12\xi}$ is close to zero. For
example, the asymptotics of $\stackrel{1}{\mu}$ as
$k\to k_0$ and $\theta\to 0$ is discontinuous:
\begin{eqnarray*}
\Big[\stackrel{1}{\mu}\vert_{k\to
k_0}\Big]\vert_{\theta\to 0}=
\bigg({|k-k_0|\over12i((k-k_0)^2-\theta^2)}
-{\bar\theta\over\theta}{k-k_0\over12i((k-k_0)^2-\theta^2)}\bigg)
\times\\
f(-\bar k_0)f(k_0)\, +\,o(1).
\end{eqnarray*}
This shows that we need  new scaling of the parameter
$\theta$. The scaled control parameter is: $$
v=t^{1/3}\,{\theta\over \sqrt{12}}. $$
\par
The external expansion constructed  in sec. 4.1
becomes discontinuous at $\theta=0$ and the internal
expansions constructed in sec. 4.1 become singular at
the point $\theta=0$ and lose their asymptotic
properties. Therefore here we must change the internal
variable for the internal  asymptotic expansion: \bb
p=t^{1/3}(k-k_0). \label{p} \ee
\par
In this section  we construct a formal  asymptotic
solution of the problem  (\ref{bar d1}) with respect
to $mod(O(t^{-1}))$ when $|\theta|\ll1$ uniform with
respect to $k\in\Complex$. The result is formulated
in the end  of this section.
\par
To  construct the uniform asymptotic solution we  need the external
and internal asymptotics outside and inside a small  neighborhood of
the point $k_0$ respectively.
\par
\begin{lemma}
Let the system of the equations (\ref{bar d1})be have
no the homogeneous nontrivial solutions, $F(k)\in
C^2\cap L_1$ and the parameters $\xi$ and $\eta$
satisfy  the inequality $|12\xi+\eta^2|\ll1$, then:
\newline $\quad$ when $|k-k_0|t^{1/3}\gg1$ the formal
asymptotic solution of the system (\ref{bar d1}) with
respect to $mod(O(t^{-2/3}/|k-k_0|)+O(t^{-1}))$ has
the form: $$
\tilde\mu=1\,+\,t^{-1}\stackrel{1}{m}(k,\xi,\eta), $$
$$ \tilde\nu=t^{-2/3}\stackrel{1}{n}_0+
t^{-1}\big(\stackrel{1}{n}_1\exp(itS)+\stackrel{2}{n}_0\big)\,
$$ the function $\stackrel{1}{m}$ is defined by
(\ref{m1}), the functions $\stackrel{1}{n}_0$ and
$\stackrel{2}{n}_2$ are defined by (\ref{n01}) and
(\ref{n02}) respectively, the function
$\stackrel{1}{n}_1$ is defined  by (\ref{ex-n-1});
\newline
$\quad$ when $|k-k_0|\ll1$ the asymptotic solution of the system
(\ref{bar d1})with respect to $mod(O(t^{-2/3}|k-k_0|+O(t^{-1}))$ has
the form: $$ \tilde\mu=1+t^{-2/3}\stackrel{1}{\cal
M}+t^{-1}\stackrel{2}{\cal M}, $$ $$
\tilde\nu=(t^{-1/3}\stackrel{1}{\cal N}+t^{-2/3}\stackrel{2}{\cal
N}+ t^{-1}\stackrel{3}{\cal N})\exp(-itS); $$ the functions
$\stackrel{j}{\cal M}$, $j=1,2$ and $\stackrel{j}{\cal N}$,
$j=1,2,3$ are defined by  (\ref{M01}), and  (\ref{N123})
respectively.
\end{lemma}
\par
{\bf Proof.} Let  us find the internal formal asymptotic expansion
for the solution of the system  (\ref{bar d1}) in the form: \bb
{\cal M}^{in}=1+t^{-2/3}\stackrel{1}{\cal M}+
t^{-1}\stackrel{2}{\cal M}+\dots, \label{fas-m0} \ee \bb {\cal
N}^{in}=(t^{-1/3}\stackrel{1}{\cal N}+ t^{-2/3}\stackrel{2}{\cal N}+
t^{-1}\stackrel{3}{\cal N}+\dots)\exp(-itS). \label{fas-n0} \ee
\par
Change the variable $k$ into the variable $p$ in the
system (\ref{bar d1}). The phase function of the
exponent is depend on the variable $p$ as: $$
tS\equiv\o(p)\equiv4(p^3+\bar p^3)-v^2(p+\bar p), $$
where $v=t^{1/3}\theta/\sqrt{12}$. Substitute
(\ref{fas-m0}) and (\ref{fas-n0}) into the system
(\ref{bar d1}), equate the coefficients with equal
power of $t$. As a result we obtain a sequence of
equation for the  coefficients of the expansions
(\ref{fas-m0}) and (\ref{fas-n0}).
\begin{eqnarray}
\pt_p
\stackrel{1}{\cal N}-i(12p^2-v^2)\stackrel{1}{\cal N}=
\sgn[\Re(-\bar p)]f(-\bar k_0), \qquad\quad\label{in-N01} \\
\pt_p \stackrel{2}{\cal N}-i(12p^2-v^2)\stackrel{2}{\cal N}=
-\sgn[-\Re(\bar p)]\big(f_{01}^{(1)}(-\bar k_0)\bar p+ \nonumber
\\
+ f^{(1)}_{10}(-\bar k)p\big) -f^{(2)}_{01}(-\bar k_0)\bar
p-f^{(2)}_{10}(-\bar k_0)p. \label{in-N02}
\\
\pt_{\bar p} \stackrel{1}{\cal M}= \sgn[-\Re(\bar
p)]f(k_0)\stackrel{1}{\cal N},\qquad\qquad \qquad\qquad\qquad
\label{in-M01} \\
\pt_p \stackrel{3}{\cal
N}-i(12p^2-v^2)\stackrel{3}{\cal
N}=\qquad\qquad\qquad\qquad\qquad\qquad\nonumber
\\ - {1\over2}\bigg[\sgn\big[-\Re(-\bar
p)\big]f^{(1)}_{20}(-\bar k_0) + f^{(2)}_{20}(-\bar
k_0)\bigg]p^2 \nonumber
\\
-{1\over2}\bigg[\sgn\big[-\Re(-\bar p)\big]f^{(1)}_{02}(-\bar k_0)+
f^{(2)}_{02}(-\bar k_0)\bigg]\bar p^2\nonumber
\\
-\bigg[\sgn\big[-\Re(\bar p)\big]f^{(1)}_{11}(-\bar k_0)+
f^{(2)}_{11}(-\bar k_0)\bigg]|p|^2\nonumber
\\
-\sgn\big[-\Re(-\bar p)\big]f(-\bar k_0)\stackrel{1}{\cal M}.
\label{in-N03}
\\
\pt_{\bar p} \stackrel{2}{\cal M} =
\sgn\big[\Re(p)\big]f(k_0)\stackrel{2}{\cal N}+
\qquad\qquad\qquad\qquad\qquad\nonumber\\
\bigg[\sgn\big[-\Re(\bar p)\big]f^{(1)}_{10}(k_0)+
f^{(2)}_{10}(k_0)\bigg]\stackrel{1}{\cal N} p+ \nonumber \\
+ \Bigg[\sgn[-\Re(k)]f^{(1)}_{01}(k_0)+f^{(2)}_{01}(k_0)\bigg]
\stackrel{1}{\cal N}\bar p. \label{in-M02}
\end{eqnarray}
\par
These equations are obtained in a supposition that
$|p|t^{-1/3}\ll1$. The uniform bounded partial
solutions of the equations for the coefficients of
the expansion (\ref{fas-n0}) are obtained by using an
integral operator: \bb {\cal P}[g]
={\exp(i\o(p))\over 2i\pi}
\int\int_{\Complex}\,{dr\wedge d\bar r\over
\overline{p-r}} \exp(-i\o(r))\,g(r), \label{cal-P}
\ee where  $g(r)$ is the right-hand side of
corresponding equation.
\par
Thus, the bounded partial solutions of the equations  (\ref{in-N01})
and (\ref{in-N02}) (the functions $\stackrel{1}{\cal N}\!\!^s$ and
$\stackrel{2}{\cal N}\!\!^s$) are represented by double  integrals.
\par
Bounded partial solutions of equations for the coefficients of the
expansion (\ref{fas-m0}) are constructed by using  the Cauchy-Green
formula.
\par
The formula for the partial solution of the equation (\ref{in-M01})
has the form: \bb \stackrel{1}{\cal M}\!\!^s(p,\xi,\eta)=
f(k_0)f(-\bar k_0)\big(J_1(p,v^2)-2J_{1}^{-+}-2J_1^{+-}\big).
\label{cal-M1s} \ee Here $$ J_1=
{1\over2i\pi}\int\int_{\Complex}{dn\wedge d\bar n\over p-n}
{\exp(i\o(n))\over2i\pi} \int\int_{\Complex}{dr\wedge d\bar r\over
\overline{n-r}} \exp(-i\o(r)). $$ $$ J_1^{\mp\pm}=
{1\over2i\pi}\int\int_{\O^{\mp}}{dn\wedge d\bar n\over p-n}
{\exp(i\o(n))\over2i\pi} \int\int_{\O^{\pm}}{dr\wedge d\bar r\over
\overline{n-r}}\exp(-i\o(r)). $$
\par
In the Appendix the integrals $J_1$ and $J_1^{-+}$
are reduced  into double integrals. By the similar
way we may represent the integral $J_1^{+-}$ as the
sum of double integrals. Therefore we may represent
$\stackrel{1}{\cal M}\!\!^s$ as sum of double
integrals also. Corresponded formula is very  large
and we don't write it here. But  we will use this
formula to  evaluate an asymptotic behaviour of
$\stackrel{1}{\cal M}\!\!^s$ as $|p|\to\infty$.
\par
The bounded partial solution of the equation (\ref{in-M02}) when
$p\in\Complex$ has to be built similarly . We give following
statement about $\stackrel{2}{\cal M}$.
\par
\begin{lemma}
Continuous partial solution of the equation for $\stackrel{2}{\cal
M}$ exists. This solution is uniformly bounded when $p\in\Complex$.
\end{lemma}
\par
{\bf Sketch of proof.} The partial solution of the
equation for $\stackrel{2}{\cal M}$ has to be
obtained as a result  of using of the operator: \bb
\pt_{\bar p}^{-1}\big[g\big]=\stackrel{2}{\cal
M}\!\!^s\equiv
{1\over2i\pi}\int\int_{\Complex}{dn\wedge d \bar
n\over p-n}g(n) \label{cal-M2s} \ee to the right-hand
side of the equation. The integrals, which are
obtained, are continuous by virtue of continuity of
the integral operator with respect to the parameter
$p$. The  boundedness  of these integrals with
respect to $p$ when $p\in\Complex$ may be obtained by
using the asymptotic behaviour of the right-hand side
terms of the equation for $\stackrel{2}{\cal M}$ as
$|p|\to\infty$.
\par
Let us to construct the external expansion in the form: \bb
m=1+t^{-1}\stackrel{1}{m}_0(k,v)\,+\,
t^{-5/3}(\stackrel{2}{m}_1\exp(itS)+\stackrel{2}{m}_0)+\dots;
\label{ex-m} \ee
\begin{eqnarray}
n &=
(t^{-1}\stackrel{1}{n}_1(k,v)+t^{-5/3}\stackrel{2}{n}_1(k,v)+t^{-2}
\stackrel{3}{n}_1\dots) \exp(itS)\, \nonumber
\\
&+
t^{-2/3}\stackrel{1}{n}_0(k,v)\,+\,t^{-1}\stackrel{2}{n}_0(k,v)+\dots.
\label{ex-n}
\end{eqnarray}
\par
Substitute (\ref{ex-m}) and (\ref{ex-n}) into
(\ref{bar d1}). As a result we get the equations for
coefficients of the expansions: \bb
\stackrel{1}{n}_1=\sgn\big[\Re(-\bar k)\big]{f(-\bar
k)\over 12i(k-k_0)^2};\quad
\stackrel{2}{n}_1=-\sgn\big[\Re(-\bar
k)\big]{v^2f(-\bar k)\over 144i(k-k_0)^4};
\label{ex-n-1} \ee \bb \pt_{\bar k}\stackrel{1}{m}=
{f(k)f(-\bar k)\over12i(k-k_0)^2}; \label{ex-m-equat}
\ee \bb \pt_k\stackrel{1}{n}_0(k,\xi,\eta)=0;\quad
\pt_k\stackrel{2}{n}_0(k,\xi,\eta)=0; \label{ex-n10}
\ee
\begin{eqnarray} \pt_{\bar
k}\stackrel{2}{m}_0={f(k)f(-\bar
k)v^2\over144i(k-k_0)^4};\\
\stackrel{2}{m}_1(k,\xi,\eta)= \sgn\big[\Re(-\bar
k)\big]f(k) {\stackrel{1}{n}_0(k,\xi,\eta)\over
12i\bar (k-k_))^2}; \label{ex-m-20}
\end{eqnarray}
\begin{eqnarray}
\stackrel{3}{n}_1(k,\xi,\eta)= {1\over
12i(k-k_0)^2}\Bigg(\sgn\big[\Re(-\bar k)\big]f(-\bar k)
\stackrel{1}{m}_0 \nonumber \\
+{2f(-\bar k)\over12i(k-k_0)^3}\sgn[\Re(-\bar k)]
{f_{10}^{(1)}(-\bar k)\sgn[\Re(-\bar k)]+f^{(2)}_{10}(-\bar k)\over
12i(k-k_0)^2}\Bigg). \label{ex-n-2}
\end{eqnarray}
\par
The formulas  (\ref{ex-n-1}) and (\ref{ex-n-2}) define the functions
$\stackrel{1}{n}_1$, $\stackrel{2}{n}_1$ and $\stackrel{3}{n}_1$.
Using the formulas (\ref{ex-n10}) we can see, that the functions
$\stackrel{1}{n}_0$ $\stackrel{2}{n}_0$ are analytic of variable
$\bar k$. The obvious form of this dependence is defined by two
conditions. First one is the continuity of the asymptotics and
second one is matching condition for the external and internal
asymptotic expansions.
\par
One can see, that the sufficient condition for the
continuity of the coefficient of the asymptotics
(\ref{ex-n}) as $t^{-1}$ with respect to $k$ is an
addition into $\stackrel{2}{n}_0$ of the term: \bb
\stackrel{2}{n'}\!_0={1\over\pi i}
{1\over12i\overline{(k-k_0)}^2}\int^{i\infty}_{-i\infty}
{d\l\over\overline{k-\l}}f(-\bar l). \label{n'} \ee
\par
It follows from the formulas (\ref{ex-m-20}), that the coefficient
of the asymptotics (\ref{ex-m}) as $t^{-5/3}$ is defined after
evaluating of $\stackrel{1}{n}_0$, i.e.  after matching of the
coefficients as $t^{-2/3}$ of the external and internal expansions.
\par
Consider the problem  for defining of $\stackrel{1}{m}$ in detail.
This function satisfies the boundary  condition: $$
\stackrel{1}{m}_0\vert_{|k|\to\infty}=0. $$
\par
The external solution is valid outside of small
neighborhoods of the points $k_j$, $j=0,1,2.$
Therefore  solutions of the equation
(\ref{ex-m-equat}), which are smooth, bounded and
decreased as $|k|\to\infty$:
\begin{eqnarray}
\stackrel{1}{m}_0\!\!^s(k,\theta)=
{1\over2i\pi}\int\int_{\Complex}{dr\wedge d\bar r \over k-r}{f(-\bar
r)f(r)-f(-\bar k_0)f(k_0)\over12i(r-k_0)^2}+ \,\nonumber\\ +\,
{\overline{(k-k_0)}f(k_0)f(-\bar k_0) \over 12i(k-k_0)^2} \label{m1s}
\end{eqnarray}
are defined to within an analytic function with
respect to variable $k$, which has poles when
$t^{1/3}|k-k_0|\ll1$. The full definition of
$\stackrel{1}{m}_0$ will be done by matching  of the
external and internal expansions.
\par
For matching process we need asymptotic behaviour of
the partial solutions of equations
(\ref{in-N01})-(\ref{in-N03}) as $|p|\to\infty$.
Constructing of these asymptotics is reduced to
evaluating of an integrals with weak  singularity in
the integrand. Evaluating of these integrals is done
in the Appendix. Here we write  the asymptotic
behaviour of partial solutions of equations
(\ref{in-N01})-(\ref{in-N03}) as $|p|\to\infty$.
\begin{eqnarray*}
\stackrel{1}{\cal N}\!^s(p,\xi,\eta) = {1\over \bar
p}\exp(i\o(p))f(-\bar
k_0)\big[\phi^+_{00}(v^2)-\phi^-_{00}(v^2)\big]+
\\
+ {1\over \bar p^2}\exp(i\o(p))f(-\bar
k)\big[\phi^+_{01}(v^2)-\phi^-_{01}(v^2)\big]+
\\
+ {1\over 12i}f(-\bar k_0)\sgn\big[\Re(-\bar p)\big] \big({1\over
p^2}-{\exp(i\o(p))\over\bar p^2}\big) \,+\,O(|v^2||p|^{-3}).
\end{eqnarray*}
\begin{eqnarray*}
\stackrel{2}{\cal N}\!^s(p,\xi,\eta) =
\bigg[{\exp(i\o(p))\over\bar
p}\big(\phi^+_{10}(v^2)-\phi^-_{10}(v^2)\big)+
\qquad
\\
\bigg({1\over p}-{\exp(i\o(p))\over\bar p}\bigg)
{\sgn[\Re(-\bar p)]\over12i}\bigg] f_{10}^{(1)}(-\bar
k_0)  +
\\
\bigg[{1\over \bar p}
\exp(i\o(p))\phi_{10}(v^2)+{1\over12ip}\bigg]f_{10}^{(2)}(-\bar k_0)
+\\
\bigg[{\exp(i\o(p))\over\bar
p}\big(\phi^+_{01}(v^2)-\phi^-_{01}(v^2)\big)+
\bigg({\bar p\over p^2}-{\exp(\o(p))\over\bar
p}\bigg) \times
\\
{\sgn\big[\Re(-\bar p)\big]\over12i}\bigg]
f_{01}^{(1)}(-\bar k_0) \,+\, \bigg[{1\over\bar
p}\exp(i\o(p))\phi_{01}(v^2)+\\
{\bar p\over12ip^2}\bigg] f^{(2)}_{01}(-\bar k_0)+
\,O(|p|^{-2})+O(|v|^2|p|^{-2}),
\end{eqnarray*}
\begin{eqnarray*}
\stackrel{1}{\cal M}\!\!^s(p,\xi,\eta)=
-f(k_0)f(-\bar k_0) \bigg[ {-\bar p\over
12ip^2}\,+\,\qquad\qquad\qquad\qquad\\
{1\over
p}\bigg(-12i\phi_{01}(v^2)\overline{\phi_{10}(v^2)} +
iv^2\phi_{00}(v^2)\overline{\phi_{00}(v^2)}\bigg)-\\
{2\over p}\bigg[
iv^2\bigg(\phi_{00}^+(v^2)\psi_{00}^-(v^2)+
\phi_{00}^-(v^2)\psi_{00}^+(v^2)\bigg)+
\\
+ 12i\bigg(\phi_{01}^+(v^2)\psi_{01}^-(v^2)+
\phi_{01}^-(v^2)\psi_{01}^+(v^2)\bigg)-{1\over2}\psi_{01}\bigg]\bigg]
\,+\,\\
\qquad\qquad\qquad\qquad\qquad O(|v|^2|p|^{-2});
\end{eqnarray*}
\begin{eqnarray*}
\stackrel{2}{\cal M}\!\!^s(p,\xi,\eta)=
f(k_0)f_{10}^{(1)}(-\bar k_0){\bar p\over 12ip}
\,+\,\qquad\qquad\qquad\qquad\\
f(k_0)f_{10}^{(2)}(-\bar k_0)\bigg({\bar p\over
12ip}+{1\over12i}\bigg)
\sgn[\Re(p)]+
\\
f(k_0)f_{01}^{(2)}(-\bar k_0)\bigg({\bar p^2\over
24ip^2}-{1\over24i}\bigg)\sgn[\Re(p)]+
\\
\qquad\qquad f(k_0)f_{01}^{(1)}(-\bar k_0) {\bar
p^2\over24ip^2}+\Phi(v^2)+O(|v|^2|p|^{-1}),
\end{eqnarray*}
\begin{eqnarray*}
\stackrel{3}{\cal N}\!^s(p,\xi,\eta) =
{1\over24i}\bigg[\sgn[\Re(-\bar
p)]\big(1-\exp(i\o(p))\big)f_{20}^{(1)}(-\bar k_0)\bigg]+
\\
\bigg[\sgn[\Re(-\bar p)]\big({\bar p^2\over12ip^2}-
{\exp(i\o(p))\over24i}\big)f_{02}^{(1)}(-\bar k_0)+
{\bar p^2f_{02}^{(2)}(-\bar k_0)\over 24ip^2}\bigg]+
\\
\bigg[\sgn[\Re(-\bar p)]\big({\bar
p\over12ip}+{\exp(i\o(p))\over 12i}\big)
f_{11}^{(1)}(-\bar k_0)+{\bar
p\over12ip}f_{11}^{(2)}(-\bar k_0)\bigg] +
\\
f_0^2(-\bar k_0)f(k_0)\exp(i\o(p))\Psi(v^2)
\,+\,O(|v||p|^{-1}).
\end{eqnarray*}
The functions $\Phi(v^2)$ and $\Psi(v^2)$ are smooth  and uniformly
bounded when $v^2\in\Real$. Here we use notations: $$
\phi^{\pm}_{mn}=\int\int_{\mp\Re(r)>0} dr\wedge d\bar r r^m \bar
r^n\exp(i\o(r)), $$ $$ \psi^{\pm}_{mn}=\int\int_{\mp\Re(r)>0}
dr\wedge d\bar r r^m \bar r^n\exp(-i\o(r)), $$ $$
\phi_{mn}=\int\int_{\Complex} dr\wedge d\bar r r^m \bar
r^n\exp(i\o(r)), $$ \bb \psi_{mn}=\int\int_{\Complex} dr\wedge d\bar
r r^m \bar r^n\exp(-i\o(r)). \label{phipsi} \ee \par Evaluate an
asymptotic behaviours of the coefficients  of the external expansion
as $k\to k_0$ and $\pt_k^2 S=o(1)$.
\begin{eqnarray*}
\stackrel{1}{n}_1|_{k\to k_0}= {f(-\bar
k_0)\over12i(k-k_0)^2}\sgn[\Re(-\bar
k)]+\qquad
\\
{1\over 12i(k-k_0)} \bigg[f_{10}^{(1)}(-\bar k) + f_{01}^{(1)}(-\bar
k_0){\overline{k-k_0}\over k-k_0} +
\\
\bigg(f_{10}^{(2)}(-\bar
k_0)+f_{01}^{(2)}(-\bar k_0)
{\overline{k-k_0}\over k-k_0}\bigg)\sgn[\Re(-\bar k)]\bigg]+ \\
+ {1\over2}f_{20}^{(1)}(-\bar k)+f_{11}^{(1)}(-\bar
k_0) {\overline{k-k_0}\over
k-k_0}+{1\over2}f_{02}^{(1)}
{\overline{(k-k_0)^2}\over(k-k_0)^2}+
\\
\bigg({1\over2}f_{20}^{(2)}(-\bar k_0) +
f_{11}^{(2)}(-\bar k_0){\overline{k-k_0}\over k-k_0}+
\\
{1\over2}f_{02}^{(2)}(-\bar
k_0){\overline{(k-k_0)^2}\over (k-k_0)^2}\bigg)
\sgn[\Re(-\bar k)]+o(1).
\end{eqnarray*}
\begin{eqnarray*}
\stackrel{2}{n}_0\!\!'=
{-1\over24\pi\overline{(k-k_0)^2}}\times\qquad\qquad\qquad
\\
\bigg[\pi
if(-\bar k_0) \sgn[\Re(-\bar p)] +
V.P.\int_{-\infty}^{\infty}
{d\l f(i\l)\over\l-ik_0}\bigg] + \\
- {1\over12\pi\overline{(k-k_0)}} \bigg[\pi
if_{01}^{(1)}(-\bar k_0)+\pi i f_{10}^{(1)}(-\bar
k_0) +
\big(\pi if_{10}{(2)}(-\bar k_0)+ \\
+ \pi if_{01}^{(2)}(-\bar k_0)\big) \sgn[\Re(-\bar k_0)] +
\\+
V.P.\int_{-\infty}^{\infty}{d\l f(i\l)\over(\l-ik_0)^2}\bigg]
+ {1\over12i}\bigg[\pi if_{20}^{(1)}(-\bar k_0)+ \\
+ 2\pi if_{11}^{(1)}(-\bar k_0) +\pi if_{02}^{(1)}(-\bar
k_0)+\big(\pi if_{20}^{(2)}(-\bar k_0)+
2\pi if_{11}^{(2)}(-\bar k_0)+ \\
+ \pi if_{02}^{(2)}(-\bar k_0)\big) \sgn[\Re(-\bar k)]+
V.P.\int_{-\infty}^{\infty}{d\l f(i\l) \over(\l-ik_0)^3}\bigg]+o(1).
\end{eqnarray*}
\begin{eqnarray*}
\stackrel{1}{m}\!\!^s=
{\overline{(k-k_0)}\over12i(k-k_0)^2}f(-\bar k_0)f(k_0)\\
{\overline{k-k_0}\over 12i(k-k_0)} \bigg[\big(f_{10}^{(1)}(-\bar
k_0)\sgn[\Re(-\bar k)] +
\\
+ f_{10}^{(2)}(-\bar k_0)\big)f(k_0) + f(-\bar
k_0)\big(f_{10}^{(1)}(k_0)\sgn[\Re(k)]+ f_{10}^{(2)}(k_0)\big)\bigg]
\\
{\overline{(k-k_0)^2}\over12i(k-k_0)^2}
\bigg[f(k_0)\big(f_{01}^{(1)}(-\bar k_0)
\sgn[\Re(-\bar k)] + f_{01}^{(2)}(-\bar k_0)\big)+ \\
+ f(-\bar k_0)\big(f_{01}^{(1)}(k_0)\sgn[\Re(k)]+
f_{01}^{(2)}(k_0)\big)\bigg]+o(1).
\end{eqnarray*}
\par
Let us do matching of the  external and internal
asymptotic expansions of the function  $\tilde \nu$.
The matching condition for $\tilde \nu$ in domain
$t^{-1/3}\ll|k-k_0|\ll1$ when $|\theta|\ll1$ has the
form:
\begin{eqnarray*}
\big(t^{-2/3}\stackrel{1}{n}_0+t^{-1}\stackrel{2}{n}_0+
t^{-1}\stackrel{1}{n_1}\exp(itS)\big) -\\
\big(t^{-1/3}\stackrel{1}{\cal
N}+t^{-2/3}\stackrel{2}{\cal N}+
t^{-1}\stackrel{3}{\cal N}\big)=o(t^{-1}).
\end{eqnarray*}
Let us equate the coefficients with equal powers of
the large parameter $t$. As a result we obtain: \bb
\stackrel{1}{n}_0={1\over\overline{k-k_0}}f(-\bar k_0)
[\phi_{00}^+(v^2)-\phi_{00}^-(v^2)], \label{n01} \ee
\begin{eqnarray}
\stackrel{2}{n}_0=& {1\over\overline{k-k_0}^2}f(-\bar
k) [\phi_{01}^+(v^2)-\phi_{01}^-(v^2)]+
\qquad\qquad\qquad\qquad\qquad\nonumber\\
&{1\over\overline{k-k_0}} f_{10}^{(1)}(-\bar
k_0)[\phi_{10}^+(v^2)\phi_{10}^-(v^2)]+
{1\over\overline{k-k_0}}f_{10}^{(2)}(-\bar k_0)
\phi_{10}(v^2)+
\nonumber\\
&{1\over\overline{k-k_0}}f_{01}^{(1)}(-\bar k_0)
[\phi_{01}^+(v^2)-\phi_{01}^-(v^2)]+ \nonumber\\
&+{1\over\overline{k-k_0}}f_{01}^{(2)}(-\bar k_0)\phi_{01}(v^2),
\label{n02}
\end{eqnarray}
where $\phi$, $\psi$, $\phi^{\pm}$ Û $\psi^{\pm}$ are defined by
(\ref{phipsi});
\begin{eqnarray}
\stackrel{1}{\cal N}={\cal P}[g_1],\quad \stackrel{2}{\cal N}={\cal
P}[g_2], \nonumber\\ \stackrel{3}{\cal N}={\cal P}[g_3] f_0^2(-\bar
k_0)f(k_0)\exp(i\o(p))\Psi(v^2), \label{N123}
\end{eqnarray}
where $g_j$ are the right-hand sides of the equations for
$\stackrel{j}{\cal N}$, the operator ${\cal P}[g]$ is defined by
(\ref{cal-P}), the function $\Psi(v^2)$ is smooth and uniformly
bounded with respect to $v^2$ when $v^2\in \Real$.
\par
Matching condition for the function $\tilde \mu$ in the domain
$t^{-1/3}\ll|k-k_0|\ll1$ when $|\theta|\ll1$ has the form: $$
\big(1+t^{-1}\stackrel{1}{m}\big)-\big(1+t^{-2/3}\stackrel{1}{\cal
M}+ t^{-1}\stackrel{2}{\cal M}\big)=o(t^{-1}). $$
\par
As a result we obtain: \bb \stackrel{1}{\cal M}=\stackrel{1}{\cal
M}\!\!^s; \quad \stackrel{2}{\cal M}=\stackrel{2}{\cal
M}\!\!^s-\Phi(v^2), \label{M01} \ee where  $\stackrel{1}{\cal
M}\!\!^s$ is defined by (\ref{cal-M1s}), the function
$\stackrel{2}{\cal M}\!\!^s$ is defined by (\ref{cal-M2s}) and the
function $\Phi(v^2)$ is smooth and uniformly bounded when
$v^2\in\Real$.
\begin{eqnarray}
\stackrel{1}{m}=\stackrel{1}{m}\!\!^s \,-\,
\, {1\over k-k_0}\bigg(-12i\phi_{01}(v^2)\overline{\phi_{10}(v^2)}
+iv^2\phi_{00}(v^2)\overline{\phi_{00}(v^2)}\bigg) \nonumber\\
 -{2\over k-k_0}\bigg[
iv^2\bigg(\phi_{00}^+(v^2)\psi_{00}^-(v^2)+
\phi_{00}^-(v^2)\psi_{00}^+(v^2)\bigg)+
\nonumber\\
+ 12i\bigg(\phi_{01}^+(v^2)\psi_{01}^-(v^2)+
\phi_{01}^-(v^2)\psi_{01}^+(v^2)\bigg)-{1\over2}\psi_{01}\bigg].
\label{m1}
\end{eqnarray}
Here $\stackrel{1}{m}\!^s$ is defined by (\ref{m1s}) and the
functions $\psi^{\pm}$, $\phi^{\pm}$, $\phi$ and $\psi$ are defined
by formulas (\ref{phipsi}).
\par
Thus we have matched internal and external expansions of $\tilde\mu$
and $\tilde\nu$. The lemma is proved.
\par
These expansions are ununiform with respect to $k$.
Now we can construct uniform asymptotic expansion with
respect to $k\in\Complex$. Following the matching
method the uniform expansion is: \bb \left(
\begin{array}{c}
\hat \mu\\ \hat\nu
\end{array}
\right)= \left(
\begin{array}{c}
\tilde m\\ \tilde n
\end{array}
\right) \,+\, \left(
\begin{array}{c}
\tilde {\cal M}_1\\ \tilde {\cal N}_1
\end{array}
\right) \,-\, A_{1,k} \left(
\begin{array}{c}
\tilde {\cal M}_1\\ \tilde {\cal N}_1
\end{array}
\right). \label{matching3} \ee
\par
Here the operator $A_{n,k}$ processes on the function
$\tilde {\cal M}_1$ in the formula for ${\cal M}_1$
by followed manner. One must change the variable $p$
into the variable $k$ using the formula (\ref{p}) and
write all terms of the asymptotic expansion with
respect to $t$ with the powers are equal to $-m$,
where $0\le m\le n$. For example, one can obtain for
the function $\tilde {\cal M}(p,\xi,\eta,t)$:
\begin{eqnarray*}
A_{1,k}[\tilde {\cal M}(p, \xi,\eta,t)]= 1+ t^{-1}
\bigg(-f(k_0)f(-\bar k_0) \bigg[ {-\overline{k-k_0}\over
12i(k-k_0)^2}\,+\,
\\
+{1\over k-k_0}\bigg(-12i\phi_{01}(v^2)\overline{\phi_{10}(v^2)} +
iv^2\phi_{00}(v^2)\overline{\phi_{00}(v^2)}\bigg)
\\
-{2\over k-k_0}\bigg[ iv^2\bigg(\phi_{00}^+(v^2)\psi_{00}^-(v^2)+
\phi_{00}^-(v^2)\psi_{00}^+(v^2)\bigg)+
\\
+ 12i\bigg(\phi_{01}^+(v^2)\psi_{01}^-(v^2)+
\phi_{01}^-(v^2)\psi_{01}^+(v^2)\bigg)-{1\over2}\psi_{01}\bigg]\bigg]
\\
f(k_0)f_{10}^{(1)}(-\bar k_0){\overline{k-k_0}\over
12i(k-k_0)} \,+\,
\\
f(k_0)f_{10}^{(2)}(-\bar
k_0)\bigg({\overline{k-k_0}\over
12i(k-k_0)}+ {1\over12i}\bigg)\sgn[\Re(k)]+ \\
f(k_0)f_{01}^{(1)}(-\bar k_0)
{\overline{k-k_0}^2\over24i(k-k_0)^2}+
\\
+ f(k_0)f_{01}^{(2)}(-\bar k_0)\bigg({\overline{k-k_0}^2\over
24i(k-k_0)^2}-{1\over24i}\bigg)\sgn[\Re(k)]\bigg),
\end{eqnarray*}
\par
The obviously formula for  $A_{1,k}[\tilde{\cal N}]$ is more large
and doesn't shown here.

\begin{theorem}
The formula (\ref{matching3}) gives the asymptotics solution of the
problem (\ref{bar d1}) with respect to $mod(O(t^{-1}))$ as
$t\to\infty$. This asymptotic solution is uniform with respect  to
$k\in\Complex$ and $|\theta|\gg1$.
\end{theorem}

\section{Justification of the asymptotics of the solution of
$\bar D$-problem}

\par
In this section we prove that the remainder of the
asymptotics has order by $t^{-4/3}$ uniformly with
respect to $k\in\Complex$ and this remainder has to
be differentiable with respect to $x$. We call by the
remainder of the asymptotics the difference between
solution of the problem (\ref{bar d1}) and
constructed asymptotic solutions (\ref{matching1})
when $\theta^2t^{-2/3}\gg1$, (\ref{matching2}) when
$-\theta^2t^{-2/3}\gg1$ and (\ref{matching3}) when
$|\theta|\ll1$. The differentiability of the
remainder will be important when we will construct an
asymptotic behaviour of solution of the equation KP-2.

\begin{theorem}
Let  $\pt^\a f(k,\bar k)\in L_1\cap C^1$ when
$|\a|\le2$ when $k$ is out of the imaginary axis and
$$ \sup_{z\in\Complex}\left\vert
\int\int_{\Complex}{dk\wedge d\bar k\over |k-z|}
|F(k)| \right\vert \,<\,2\pi, $$ then the solution of
the problem (\ref{bar d1}) is: \bb \left(
\begin{array}{c}
\mu\\
\nu
\end{array}
\right)= \left(
\begin{array}{c}
\tilde\mu\\
\tilde\nu
\end{array}
\right)+O(t^{_4/3}), \label{sol.bar d } \ee when $k\in \Complex, \,
\xi,\eta\in \Real$. The remainder of the asymptotics has to  be
differentiable with respect to $x$.
\end{theorem}
\par
{\bf The proof.} Let us write the system of differential  equation
for the remainder. Denote the remainder in (\ref{sol.bar d}) by $V$.
Substitute (\ref{sol.bar d}) into (\ref{bar d1}). As a result we
obtain:
\bb \left(\begin{array}{cc} \pt_{\bar k}&0\\
0&\pt_{k}
\end{array}\right)
V= \left(\begin{array}{cc}
0&F(-\bar k)\exp(itS)\\
F(k)\exp(-itS)&0
\end{array}\right)
V+f, \label{rem1} \ee \bb V|_{|k|\to\infty}=0.
\label{rem2} \ee We denote  by the vector $f$ the
residual which originates in (\ref{bar d1}) when we
substitute the column into this equation  $(\tilde
\mu,\tilde\nu)^T$: $$ f_1=-\pt_{\bar k}\tilde
\mu\,+\,F(-\bar k)\exp(itS)\tilde\nu, $$ $$
f_2=-\pt_k\tilde\nu\,+ \, F(k)\exp(-itS)\tilde\mu. $$
\par
Let us denote by $X$ the space of bounded and continuous with
respect to $k$ vector-functions with the  norm: $$
||W||=\sup_{k\in\Complex, (\xi,\eta)\in\Real^2} |W_1| \,+\,
\sup_{k\in\Complex, (\xi,\eta)\in\Real^2} |W_2|. $$
\par
Consider a system of integral equations instead of
the problem (\ref{rem1}), (\ref{rem2}): \bb V=G[F]V+H,
\label{rem3} \ee where $G[F]$ is the integral
operator:
\begin{eqnarray*}
G[F]V= \int\int_{m\in\Complex}dm\wedge d\bar m
\times\qquad\qquad\qquad\qquad
\\
\left(
\begin{array}{cc}
0 & {F(-\bar m)\over k-m}\exp(itS)\\
{F(m)\over \overline{k-m}}\exp(-itS) & 0
\end{array}
\right) V(m,\xi,\eta,t); \end{eqnarray*} $$
H=\int\int_{m\in\Complex}dm\wedge d\bar m \left(
\begin{array}{c}
{f_1(m,\xi,\eta,t)\over k-m}\\
{f_2(m,\xi,\eta,t)\over \overline{k-m}}
\end{array}
\right). $$
\par
Using obvious forms of the functions $f_{1,2}$ one can
proof, that  $$ ||H||=O(t^{-4/3}) $$ uniformly with
respect to $\xi,\eta\in\Real$ when $k\in\Complex$.
\par
The  operator $G[F]$ is contracting in  the space $X$,
therefore the solution of the integral equation
(\ref{rem3}) exists in $X$ and is evaluated by
$O(t^{-4/3})$ uniformly with respect to
$\xi,\eta\in\Real$.
\par
Show that  the remainder of the asymptotics is differentiable on
$x=t\xi$. Differentiate with respect to $x$ the system of the
equation for the remainder. Denote the derivative of the vector $V$
by $\chi$. Then we obtain: $$ \chi=G[F]\chi+\pt_x G[F]V+\pt_x H. $$
The terms $\pt_x G[F]V+\pt_x H$ may be evaluated by order
$O(t^{-4/3})$. The operator $G[F]$ is contracting, therefore one can
obtain: $||\chi||=O(t^{-4/3})$. The theorem is proved.
\par

\section{Solution of the equation KP-2}

\par
An asymptotics of the solution of the problem (\ref{bar d}) as
$t\to\infty$ may be written as: \bb \left(
\begin{array}{c}
\phi\\
\psi
\end{array}
\right) = \left(
\begin{array}{c}
\tilde\mu\\
\tilde\nu
\end{array}
\right)(k,\bar k,\xi,\eta,t)+ \left(
\begin{array}{c}
-\tilde\nu\\
\tilde\mu
\end{array}
\right)(-\bar k,-k,\xi,\eta,t)+ O(t^{-4/3}). \label{sol.bar d} \ee
\par
The second term in this formula is the solution of the problem
(\ref{bar d}) with the boundary condition: $$ \left(
\begin{array}{c}
\tilde\mu\\
\tilde\nu
\end{array}
\right)\Bigg|_{|k|\to\infty} = \left(
\begin{array}{c}
0\\
1
\end{array}
\right). $$
\par
{\bf The proof of the theorem  1.} Let us substitute
the function $\psi(k,\bar k,\xi,\eta,t)$ into the
formula for the  solution of the equation
KP-2(\ref{sol}). Differentiate  the integrand with
respect to $x$. The main terms of the integrand are
the terms which appear after  differentiating of the
exponent. The derivatives with respect to $x$ of
others factors of integrand are small because they
depend on $x$ slowly. The integrals of such terms
over the plane are evaluated by the order
$O(t^{-4/3})$ uniformly with respect to $(x,y)\in
\Real$. Rewrite the main term of the integral as the
integral over real plane: $(\k,\l)\in\Real^2,$ where
$\k=\Re(k),\quad \l=\Im(k)$. As a result we obtain:
\begin{eqnarray}
u(x,y,t)=-4\int\int_{\Real^2}\, d\k\wedge d\l\,|\k|f(\k+i\l)\times
\nonumber\\ \times\exp(it(8\k^3-24\k\l^2+2\k\xi+4\k\l\eta))\,
+\,O(t^{-4/3}). \label{as.sol kp}
\end{eqnarray}
\par
Thus the main term of the asymptotics of the solution of the Cauchy
problem  for the equation KP-2 is given by integral with fast
oscillating exponent. Let us evaluate the asymptotic behaviour of
this integral.
\par
Let  $\eta^2+12\xi>0$ and
$t^{1/3}|\eta^2+12\xi|\gg1$, then  the stationary
points of the exponent are:
$\l_{1,2}={\eta\over12}\pm\sqrt{\eta^2+12\xi},\quad
\k_{1,2}=0$. The stationary phase method (see  for
example \cite{Fedoryk}) gives: $$ u(x,y,t)=o(t^{-1}).
$$
\par
Let $\eta^2+12\xi<0$  and
$t^{1/3}|\eta^2+12\xi|\gg1$, then
$\l_{1,2}={\eta\over12},\quad
\k_{1,2}=\pm{1\over2}\sqrt{-\eta^2-12\xi}.$ The
asymptotic behaviour of the integral is:
\begin{eqnarray*}
u(x,y,t)=-4t^{-1}{\pi\over12i\sqrt{-\eta^2-12\xi}}
f_0\bigg({1\over2}\sqrt{-\eta^2-12\xi}+{i\eta\over12}\bigg)\times
\\
\times \exp\bigg(-11it\sqrt{-{y^2\over t^2}-12{x\over
t}}\bigg)+c.c.+o(t^{-1}). \end{eqnarray*}
\par
To evaluate the main term of the asymptotics of the
integral when $|12\xi+\eta^2|=o(1)$, we substitute
the scaled variables $p_1=t^{1/3}{\hbox{Re}}(k-k_0)$,
$p_2=t^{1/3}{\hbox{Im}}(k-k_0)$ and the parameter
$v^2=t^{2/3}(\eta^2+12\xi)/\sqrt{12}$ into integral
(\ref{as.sol kp}). As a result we obtain:
\begin{eqnarray*}
u(\xi,\eta,t)=4it^{-1}f(k_0)\int\int_{\Real^2}dp_1
dp_2\, \times\qquad\qquad\qquad\qquad
\\
p_1\,\exp(i(8p_1^3-2v^2p_1-24p_1p_2^2))
\,+\,o(t^{-1}).
\end{eqnarray*}
\par
Let us integrate the internal integral with respect to the parameter
$p_2$, use the even property of the  integrand with respect to
$p_1$. As a result we obtain:
\begin{eqnarray*}
u(x,y,t)=8it^{-1}\sqrt{\pi}f(i\eta/12) \bigg( \int_0^\infty
dp_1\sqrt{p_1} \cos\big(8p_1(p_1^2-8v^2)\big)+
\\+
\int_0^\infty dp_1\sqrt{p_1} \sin\big(8p_1(p_1^2-8v^2)\big)\bigg)
\,+\,o(t^{-1}).
\end{eqnarray*}
The theorem 1 is proved.

\appendix
\section{Asymptotic behaviour of double integral with weak singularity
of  the integrand}

\par
Here we obtain the asymptotic  behaviour of integrals
which are appeared when the asymptotic solution of
(\ref{bar d1}) was studied.
\par
Evaluating of the asymptotic behaviour of
one-dimensional integrals with weak singular
integrand and fast oscillated exponent was done in
\cite{Fedoryk} p.26 and  \cite{Olver} p.332. The
asymptotic behaviour of many-dimensional integrals
with fast oscillating exponent was studied in
\cite{Fedoryk}, \cite{Arn-Var-G.Z}. The asymptotic
behaviour of the Cauchy integrals with fast
oscillating exponent in one-dimensional case was
studied in \cite{Fedoryk} and in many-dimensional
case was studied in \cite{ok2}. The asymptotic
behaviour of the two-dimensional for some integrals
over all complex plane with weak singularity was
studied in \cite{ok}.

\subsection{Integrals over half-plane with general
stationary point of fast oscillated exponent}

\par
Here we study an asymptotic behaviour of  an  integral: $$
I=\int\int_{\O^+}{dn\wedge d\bar n\over l-n}\exp(-i(n^2+\bar n^2)),
$$ where $|l|\to\infty$ and  the domain $\O^+=\{\Re(l+\bar
l-i(l-\bar l))>0\}$.
\par
\begin{theorem} The asymptotic behaviour of the integral $I$ as
$|p|\to\infty$ has the form:
$$ I=-2i\pi{\exp(-i(l^2+\bar l^2))\over 2il}\,-\, {3i\pi\over 2\bar
l} \,+\,O(|l|^{-2})\quad {\hbox{ѳÛ}}\quad l\in\O^+; $$ $$
I=-{i\pi\over 2\bar l}\,+\,O(|l|^{-2})\quad {\hbox{where}}\quad
l\not\in\O^+. $$
\end{theorem}
\par
{\bf The proof.} Let us suppose that $l\in\O^+$.
Divide the domain $\O^+$ into three domains: first
one is $\O^+_1=\{\O^+\backslash\{[|n|\le
|l|/2]\cup[|n-l|<\ve]\}\}$. This domain hasn't
stationary point of phase function of the exponent
and the singularity of the integrand. Second one is
$\O^+_2=\{\O^+\backslash [|n|\ge|l|/2]\}$. This
domain contains the stationary point of the phase
function. At last, third domain  is
$\O_3^+=\{|n-l|<\ve\}$. This domain contains the
singularity of the integrand.
\par
Let us integrate by parts over $\O^+_1$. As a result
we obtain: $$ \int\int_{\O_1^+}{dn\wedge d\bar n\over
\overline{l-n}} \exp(-i(n^2+\bar n^2))= \int_{\pt
\O_1^+}{d\bar n\over\overline{l-n}} {\exp(-i(n^2+\bar
n^2))\over -2in} \,-\, $$ \bb -\,
\int\int_{\O_1^+}{dn\wedge d\bar n\over
\overline{l-n}} {\exp(-i(n^2+\bar n^2))\over 2in^2}.
\label{I-1} \ee The boundary of the domain $\O_1^+$
includes a large half-circle of radius  $R$ as
$R\to\infty$, a circle of radius $\ve$ over the point
$l=n$, a half-circle of radius  $|l|/2$ and two
segments: $[R\exp(3i\pi/4), |l|\exp(3i\pi)/2]$ and
$[|l|\exp(-i\pi/4)/2, R\exp(3i\pi)]$.
\par
Consider the integrals over boundary  of the domain
$\O_1^+$. The integral over large half-circle is
equals to zero as $R\to\infty$. The integral over
half-circle of radius  $|l|/2$ has order $|l|^{-3}$
(because of oscillations and small value of the
integrand). The integral over the circle at $n=l$
equals to  a residue of the integrand multiplied by
$2i\pi$ as  $\ve\to0$. The integral over the segments
$[R\exp(3i\pi/4),|l|\exp(3i\pi)/2]$ and
$[|l|\exp(-i\pi/4)/2,R\exp(3i\pi)]$ (let us denote
their union by $L$) is reduced to the form: $$
\int_{n\in L} {d\bar n\over -2in\overline{(l-n)}}=
{\exp(3i\pi/4)\over -2i\bar l}
\int_{|\l|\ge|l|/2}{d\l\over \bar l-\l\exp(i\pi/4)}.
$$
\par
The second term of (\ref{I-1}) has to be evaluated as by
$O(|l|^{-3})$.
\par
Let us consider  the integral over $\O_2^+$.
Represent: $$ {1\over \overline{l-n}}={1\over\bar
l}\big(1+{\bar n\over\overline{l-n}}\big). $$ Then
the integral over $\O_2^+$ has to be written as:
\begin{eqnarray*}
I_2=\int\int_{\O_2^+}{dn\wedge d\bar n\over
\overline{l-n}}\exp(-i(n^2+\bar n^2))=
\\
{1\over\bar l}\int\int_{\O_2^+}{dn\wedge d\bar n\over
\overline{l-n}} \exp(-i(n^2+\bar n^2))+
\\
{1\over\bar l}\int\int_{\O_2^+}{dn\wedge d\bar n\over
\overline{l-n}} \bar n\exp(-i(n^2+\bar n^2)).
\end{eqnarray*}
Here we integrate by  parts  the second term. After
evaluations we obtain:
\begin{eqnarray*}
I_2={1\over \bar l}\int\int_{\O^+}dn\wedge d\bar
n\exp(-i(n^2+\bar n^2))\,+\,
\\
{1\over 2i\bar l}
\int_{|l|\exp(3i\pi/4)/2}^{|l|\exp(-i\pi/4)/2}
{dn\over\overline{l-n}} \,+\,O(|l|^{-2}).
\end{eqnarray*}
\par
The integral over $\O_3^+$ as $\ve\to 0$ equals
to zero.
\par
Let us sum the obtained asymptotics: $$ I=-2i\pi{\exp(-i(l^2+\bar
l^2))\over 2il}\,+\, {\exp(i\pi/2)\over 2i\bar l}
\int_{-\infty}^{\infty}{d\l\over \bar l\exp(-i\pi/4)-\l} \,+\, $$ $$
{1\over 2i\bar l}\int_{\O^+}dn\wedge d\bar n\exp(-i(n^2+\bar n^2))
+O(|l|^{-2}). $$
\par
Thus the first statement  of the theorem is proved.
The second statement has to be proved by the same way.
\par
The theorem  is proved.

\subsection{Asymptotic behaviour of the  integral\\ with
confluent phase function}

\par
In this section  we obtain an asymptotic behaviour of an integral as
$|p|\to\infty$ \bb W^+=\int\int_{\O^+}{dr\wedge d\bar
r\over\overline{p-r}} \exp(-i\o(r)). \label{int1} \ee \par
\begin{theorem}
The asymptotic behaviour  of the integral
(\ref{int1}) where $\o(p)$ $=4(p^3+\bar
p^3)-v^2(p+\bar p)$ as $|p|\to\infty$ and
$|p\pm{v\over\sqrt{12}}|\ge0$ has the form: \bb W^+=
\left[
\begin{array}{c}
{1\over \bar p}\phi_{00}^+(v^2)+{1\over\bar p^2}\phi_{01}^+(v^2)+
{\pi i\over 12\bar p^2}+2\pi i {\exp(-i\o(p))\over 12p^2}+ \\
+O(|p|^{-3}+|v|^2|p|^{-2}), \quad {\hbox{when}}\quad p\in\O^+; \\
\quad\\
{1\over \bar p}\phi_{00}^+(v^2)+{1\over\bar p^2}\phi_{01}^+(v^2)
{\pi i\over 12\bar p^2}+ \\
+O(|p|^{-3}+|v|^2|p|^{-2}), \quad {\hbox{when}}\quad p\not\in\O^+.
\end{array}
\right. \label{as1-int1} \ee
\end{theorem}
\par
Let us prove the theorem. Represent the  integral in the form:
\begin{eqnarray*} W^+={1\over \bar p} \int\int_{\O^+}dn\wedge d\bar
n\exp(-i\o(n))\,+\\ +\, {1\over \bar p^2} \int\int_{\O^+}dn\wedge
d\bar n\exp(-i\o(n))\bar n\, +\,\\ +{v^2\over 12\bar p^2}
\int\int_{\O^+}{dn\wedge d\bar n\over \overline{p-n}}\bar
n^2\exp(-i\o(n))\,+\,
\\+
{1\over 12\bar p^2} \int\int_{\O^+}{dn\wedge d\bar n\over
\overline{p-n}} (12\bar n^2-v^2)\exp(-i\o(n)).
\end{eqnarray*}
Integrate by parts the last term. As a result we obtain:
\begin{eqnarray*}
W^+={1\over \bar p}\phi_{00}^+(v^2)+ {1\over\bar
p^2}\phi_{01}^+(v^2)+
\qquad\qquad\qquad\qquad\\+{1\over 12\bar
p^2}\int_{\pt\O^+}{d\bar n\over \overline{p-n}}
{(12\bar n^2-v^2)\exp(-i\o(n)))\over
(-i)(12n^2-v^2)}+\\ O(|p|^{-3}+|v|^2|p|^{-2}).
\end{eqnarray*}
Consider  the integral over the boundary of  the
domain $\pt\O^+$. This integral may be considered  as
a sum of integrals over imaginary axis $\Re(r)=0$,
over half-circle of radius $R$ as $R\to\infty,\quad
\Re(r)>0$ and over the circle $|p-r|=\ve$ as $\ve\to
0$ (if $p\in\O^+$). We can see that the integral over
the large half-circle tends to zero. As a result we
obtain the statement of the theorem.
\par
One more double  integral which we need is: $$
U^+=\int\int_{\O^+}{dr\wedge d\bar r\over p-r} \,\bar
r\exp(-i\o(r)). $$ Its asymptotics  can be evaluated
by the same way as the asymptotics  of the integral
(\ref{int1}) as $|p|\to\infty$ $|p\pm v|\ge0$:
\begin{eqnarray*}
U=\left[
\begin{array}{c}
{1\over \bar p}\phi^+_{10}(v^2)+2i\pi{1\over12i p}\exp(-i\o(p))+
{\pi i\over 12\bar p}+2\pi i {\exp(-i\o(p))\over 12p^2}+
\\+
O(|p|^{-2}+|v|^2|p|^{-2}), \quad {\hbox{when}}\quad p\in\O^+; \\ \quad\\
{1\over \bar p}\phi_{10}^+(v^2){\pi i\over 12\bar p}+
O(|p|^{-2}+|v|^2|p|^{-2}), \quad {\hbox{when}}\quad
p\not\in\O^+.
\end{array}
\right.
\end{eqnarray*}

\section{Reducing of four-multiply integral into double integral}

\par
This section is pure technical. Here we  show as the four-multiply
integrals have to be reduced into double integrals over half-plane
and all complex plane.

\subsection{Four-multiply integral with nondegenerate phase
of the exponent.}

\par
Let  us show that the four-multiply integral may be
written as a sum of double integrals. Change the
variable: $n-m=r$,  then the integral $J$ has the
form:
\begin{eqnarray*}
J={1\over2i\pi}\int\int_{\Complex}{dn\wedge d\bar
n\over l_j-n}\times\qquad\qquad\qquad\qquad
\\
{1\over2i\pi} \int\int_{\Complex}{dr\wedge d\bar
r\over \bar r}\exp(i(r^2+\bar r^2))
\exp(-2i(rn+\overline{rn})).
\end{eqnarray*}
Integrate by part over  $\bar n$. As  a result we
obtain:
\begin{eqnarray*}
J={-1\over4\pi^2}\lim_{R\to\infty}\int_{|n|=R}{dn
\bar n\over l_j-n} \times\qquad\qquad
\\
\int\int_{\Complex}{dr\wedge d\bar r\over \bar
r}\exp(i(r^2+\bar r^2))
\exp(-2i(rn+\overline{rn}))\,+
\\
{1\over2i\pi}\lim_{\ve\to0}\int_{|l-n|=\ve}{dn \bar
n\over l_j-n} \times\qquad\qquad
\\
{1\over2i\pi}
\int\int_{\Complex}{dr\wedge d\bar r\over \bar
r}\exp(i(r^2+\bar r^2)) \exp(-2i(rn+\overline{rn}))\,-
\\
{1\over2i\pi}\int\int_{\Complex} {dn\wedge d\bar
n\over l_j-n}\bar n\times\qquad\qquad
\\
{1\over2i\pi} \int\int_{\Complex}dr\wedge d\bar
r(-2i)\exp(i(r^2+\bar r^2))
\exp(-2i(rn+\overline{rn})).
\end{eqnarray*}
First term is equal to zero because of fast
oscillating of the integrand, second term  is equal
to residue with  sign minus  of the integrand at
$n=l$, the internal integral in the third term has to
be evaluated. Finally we obtain: $$ J=\bar
l_j{\exp(i(l_j^2+\bar l_j^2))\over
2i\pi}\int\int_{\Complex} {dn\wedge d\bar n\over
\overline{l_j-m}}\exp(-i(n^2+\bar n^2))\,-\,
\exp(i(l_j^2+\bar l_j^2)). $$

\subsection{Four-multiply integral over different half-planes with
nondegenerate phase function}

\par
Let us consider an integral:
\begin{eqnarray*}
J_{-+}=\int\int_{\O^{-}}{dn\wedge d\bar n \over
l-n}{\exp(i(n^2+\bar n^2)) \over
2i\pi}\times
\\
\int\int_{\O^{+}}{dm\wedge d\bar m \over
\overline{n-m}} \exp(-i(m^2+\bar m^2)).
\end{eqnarray*}
\par
A partially integrating over  $\bar n$ gives:
\begin{eqnarray*}
J_{-+}=\int_{\pt \O_-}{dn\over l-n}\bar
n{\exp(i(n^2+\bar n^2))\over2i\pi}\times\qquad\qquad
\\
\int\int_{\O^{+}}{dm\wedge d\bar m \over
\overline{n-m}} \exp(-i(m^2+\bar m^2))\,-\,
\\
\int\int_{\O^{-}}{dn\wedge d\bar n \over l-n}\bar n
{\pt\over \pt \bar n} {\exp(i(n^2+\bar n^2))\over
2i\pi}\times\qquad\qquad
\\
\int\int_{\O^{+}} {dm\wedge d\bar m \over
\overline{n-m}}\exp(-i(m^2+\bar m^2)).
\end{eqnarray*}
The derivative is:
\begin{eqnarray*}
J_{-+}=\int_{\pt
\O^{-}}{dn\over l-n}{\bar n\exp(i(n^2+\bar n^2))\over
2i\pi}\times\qquad\qquad
\\
 \int\int_{\O^{+}}{dm\wedge d\bar
m\over\overline{n-m}} \exp(-i(m^2+\bar m^2))\,-\,
\\
{3\over4}\int_{\pt \O^{-}}{dn\over
l-n}\exp(i(n^2+\bar n^2)).
\end{eqnarray*}
\par
Let us consider integral  over part of the boundary in first term :
\begin{eqnarray*}
{\cal I}=\lim_{R\to\infty} \, \int_{R\exp(-i\pi/4)}^{R\exp(3i\pi/4)}
{dn\over l-n}{\bar n\exp(i(n^2+\bar n^2))\over 2i\pi} \times
\\
\int\int_{\O^{+}}{dm\wedge d\bar m\over\overline{n-m}}
\exp(-i(m^2+\bar m^2)).
\end{eqnarray*}
Change the order of integration over $m,\bar m$ and  $n$. The
internal integral over  $n$ has to be evaluate: $$
\lim_{R\to\infty}\int_{R\exp(-i\pi/4)}^{R\exp(3i\pi/4)} dn {\bar
n\over (l-n)\overline{(n-m)}}= \left[
\begin{array}{c}
{1\over 2}+{\bar m\over il-\bar m}, \quad {\hbox{when}}\quad l\in\O^+; \\
-1/2\quad {\hbox{when}}\quad l\in\O^+.
\end{array}
\right. $$
\par
In last expression of $J_{-+}$ the integrals over the  large half of
circle $|n|=R$ as $R\to\infty$ are equal to zero.
\par
Final formulas for $J_{-+}$ have to be written as: $$ J_{-+}= \left[
\begin{array}{c}
{-5\over4}i\pi+il\,\int\int_{\O^+}dm\wedge d\bar m
{\exp(-i(m^2+\bar m^2)) \over il-\bar m},
\quad{\hbox{when}}\quad l\in\O^+; \\ \quad \\
{-5\over 4}i\pi \,+\, {3\over2}i\pi\exp(i)l^2+\bar
l^2)) - \bar l\exp(i(l^2+\bar
l^2))\times\\
\int\int_{\O^+}dm\wedge d\bar m {\exp(-i(m^2+\bar
m^2))\over\overline{l-m}},\,\,{\hbox{when}}\,\,\,
l\in\O^-. \end{array} \right. $$

\subsection{Four-multiply integral with confluent \\
phase function}

\par
Let us reduce an four-multiply integral $$
J_1=\int\int_{\Complex}{dn\wedge d\bar n \over p-n}\exp(i\o(n))
\int\int_{\Complex}{dm\wedge d\bar
m\over\overline{n-m}}\exp(-i\o(m)) $$ into sum of double integrals.
Denote: $$ V(n,v^2)={\exp(i\o(n))\over2i\pi}\int\int_{\Complex}
{dr\wedge d\bar r\over\overline{n-r}}\exp(-i\o(r)). $$ The function
$V(p,v^2)$ has to be written as: $$ {\pt V\over\pt \bar n}=12i\bar
n\exp(i\o(n))\phi_{00}(v^2)\,+\,12i\exp(i\o(n))\phi_{01}(v^2). $$ We
can obtain this formula by changing variable in the integrand
$\rho=n-r$ and differentiating  the obtained  expression of
$V(n,v^2)$ with respect to $\bar n$ and, finally changing the
variable back: $r=n-\rho$.
\par
Partial differentiating of the double integral over $\bar n$ and $n$
gives: $$ \int\int_{\Complex}{dn\wedge d\bar n\over p-n}V(n,v^2)=
\lim_{R\to\infty}\int_{|n|=R}{dn\,\bar n\,V(n,v^2)\over p-n}\,+\, $$
\bb + \lim_{\ve\to 0}\int_{|p-n|=\ve}{dn\,\bar n\over
p-n}V(n,v^2)\,-\, \int\int_{\Complex}{dn\wedge d \bar n\over
p-n}\bar n{\pt V(n,v^2)\over\pt\bar n}. \label{by-parts} \ee
Asymptotic behaviour of $V(n,v^2)$ as $|n|\to\infty$ was obtained
above. Using this asymptotics for the first part of the formula
(\ref{by-parts}) gives zero. Second term of the right-hand side of
(\ref{by-parts}) gives the residue of the integrand at $p$
multiplied by $2i\pi$. Then we obtain: $$ J_1(p,v^2)=\bar
p\,V(p,v^2)\,-\,\phi_{00}(v^2)\exp(i\o(p))\,-\, $$ $$
-\,12i\phi_{01}(v^2){1\over2i\pi} \int\int_{\Complex}{dn\wedge d\bar
n\over p-r}\bar n\exp(i\o(n)) \,+\, $$ $$ \,+\,
iv^2\phi_{00}(v^2){1\over2i\pi}\int\int_{\Complex}{dn\wedge d\bar
n\over p-r} \exp(i\o(n)). $$

\par
On the same way one can evaluate an integral $$
J_1^{-+}=\int\int_{\O^-}{dn\wedge d\bar n\over
p-n}\exp(i\o(n)) \int\int_{\O^+}{dm\wedge d\bar
m\over \overline{n-m}}\exp(-i\o(m)). $$
\par
The difference between the integral $J_1$ and the
integral $J_1^{-+}$ consists  in the addition terms
over boundaries of the domains $\O^+$ and  $\O^-$ in
the result. The finally forms are:
\newline when $p\in\O^+$:
\begin{eqnarray*}
J_1^{-+}=-\phi^+_{00}(v^2)\exp(i\o(p))+
\\
+\bar p\exp(i\o(p))\int\int_{\O^+}{dm\wedge d\bar m\over
\overline{p-m}}\exp(-i\o(m))
\\
{1\over2}\int\int_{\O^-}{dn\wedge d\bar n\over p-n} \bar
n\exp(i\o(n))+
\\+
iv^2\phi_{00}^+(v^2){1\over 2i\pi}
\int\int_{\O^-}{dn\wedge d\bar n\over p-n}\exp(i\o(n))\\
-12i\phi_{01}^+(v^2){1\over2i\pi} \int\int_{\O^-}{dn\wedge d\bar
n\over p-n} \bar n\exp(i\o(n));
\end{eqnarray*}
when $p\not\in\O^+$: $$
J_1^{-+}=-\phi^+_{00}(v^2)+p\int\int_{\O^+}{dm\wedge
d\bar m\over p+\bar m}\exp(-i\o(m)) $$
\begin{eqnarray*}
-{1\over2}\int\int_{\O^-}{dn\wedge d\bar n\over
p-n}\bar n\exp(i\o(n))\\
iv^2\phi_{00}^+(v^2){1\over2i\pi}
\int\int_{\O^-}{dn\wedge d\bar n\over p-n}\exp(i\o(n))\\
-12i\phi_{01}^+(v^2){1\over2i\pi} \int\int_{\O^-}{dn\wedge d\bar
n\over p-n} \bar n\exp(i\o(n)).
\end{eqnarray*}

\medskip
\centerline{{\bf\large Acknowledgements.}} \par I am grateful to
S.G. Glebov, A.M. Il'in, L.A. Kalyakin and M.M. Shakir'yanov for
stimulated discussions.

\end{document}